\documentclass[prb,twocolumn, superscriptaddress,amsmath,amssymb,floatfix]{revtex4-1}
\usepackage{graphicx}
\usepackage{dcolumn}
\usepackage{bm}
\usepackage{textcomp}
\usepackage{gensymb}
\usepackage{multirow}
\usepackage[none]{hyphenat}
\usepackage{epstopdf}
\usepackage{float}
\usepackage{tabularx}
\usepackage{color}
\usepackage{siunitx}
\usepackage{xr}
\usepackage{inputenc}

\usepackage{amsmath} 
\usepackage{mathtools}
\usepackage[normalem]{ulem}
\usepackage{natbib}
\usepackage[caption=false]{subfig}
\usepackage{afterpage}
\usepackage{enumerate}
\usepackage{hyperref}
\usepackage{color,soul}
\usepackage[all]{hypcap}
\makeatletter
\newcommand*{\rom}[1]{\expandafter\@slowromancap\romannumeral #1@}
\makeatother

\renewcommand{\vec}[1]{\bm{#1}}

\newcommand{\beq}{\begin{equation}}
\newcommand{\eeq}{\end{equation}}
\newcommand{\kk}{\mathbf{k}}

\begin{document}

\title{Inherited topological superconductivity in two-dimensional Dirac semimetals}

\author{Chiu Fan Bowen Lo}
\affiliation{Department of Physics, Columbia University, New York, NY 10027, USA}

\author{Hoi Chun Po}
\email[Correspondence e-mail address: ]{hcpo@ust.hk}
\affiliation{Department of Physics, Massachusetts Institute of Technology, Boston, MA 02139, USA}
\affiliation{Department of Physics, Hong Kong University of Science and Technology, Clear Water Bay, 999077 Hong Kong, China}

\author{Andriy H. Nevidomskyy}
\email[Correspondence e-mail address: ]{nevidomskyy@rice.edu}
\affiliation{Department of Physics and Astronomy \& Rice Center for Quantum Materials, Rice University, Houston, TX 77005, USA}

\date{\today}

\begin{abstract}
Under what conditions does a superconductor inherit topologically protected nodes from its parent normal state? In the context of Weyl semimetals with broken time-reversal symmetry, the pairing order parameter is classified by monopole harmonics and necessarily nodal [Li and Haldane, Phys. Rev. Lett., 120, 067003 (2018)]. Here, we show that a similar conclusion could also apply to 2D Dirac semimetals, although the conditions for the existence of nodes are more complex, depending on the pairing matrix structure in the valley and sublattice space. We analytically and numerically analyze the Bogoliubov-de-Gennes quasi-particle spectra for Dirac systems based on the monolayer as well as twisted bilayer graphene. We find that in the cases of intra-valley intra-sublattice pairing, and inter-valley inter-sublattice pairing, the point nodes in the BdG spectrum (which are inherited from the Dirac cone in the normal state) are protected by a 1D winding number. The nodal structure of the superconductivity is confirmed using tight-binding models of monolayer and twisted bilayer graphene. Notably, the BdG spectrum is nodal even with a momentum-independent ``bare" pairing, which, however, acquires momentum-dependence and point nodes upon projection to the Bloch states on the topologically nontrivial Fermi surface, similar in spirit to the Li--Haldane monopole superconductor and the Fu--Kane proximity-induced superconductor on the surface of a topological insulator.
\end{abstract}

\maketitle

\section{Introduction}

Since the discovery of topological insulators (TI) more than a decade ago~\cite{Zhang2009,Hasan2010TI,Xia2009,Chen178,Qi2011}, there is a growing body of examples of symmetry protected topological phases of matter, classified in the non-interacting limit by the discrete symmetries of the Hamiltonian~\cite{Schnyder2008,kitaev2009,10fold-way}, including crystalline symmetries~\cite{Fu-TCI}.
Included in this classification are topological superconductors, characterized by the particle-hole (charge conjugation) symmetry of the Bogoliubov-de-Gennes (BdG) Hamiltonian. In the original, strict sense of the term, topological superconductivity refers to fully gapped phases, such as $(p_x+ip_y)$ superconductor (class A) \cite{qi_hughes_raghu_zhang_2009,read_green_2000} in two dimensions or the B-phase of $^{3}$He (class DIII) in 3D \cite{anderson_morel_1961,balian_werthamer_1963, leggett_1975,Volovik2010}. In a broader sense, which we shall adopt for the rest of this article, topological superconductors also include the gapless phases, where the nodes of the superconducting gap (or more precisely, the nodes of the BdG quasi-particle spectrum) are topologically protected~\cite{sato_ando2017}; i.e. the presence of such gap nodes is not accidental but is necessitated by the underlying topology of the normal state, even if one considers a featureless $s$-wave pairing in the microscopic Hamiltonian. The nodes appear upon projecting this ``bare" pairing onto the Fermi surface, morally similar to how the momentum dependence of the pairing develops in the Fu--Kane mechanism of proximity induced topological $p$-wave superconductivity~\cite{fu_kane2008}. When and how does the superconducting state inherit the normal state topology? The most general answer to this question is not presently known, although several examples of concrete constructions exist in 2D and 3D (doped) semimetals, which we summarize below.

In a 2D tight-binding Haldane model of graphene with complex next-nearest neighbour interactions, it was shown by Murakami and Nagaosa~\cite{murakami_nagaosa2003} that the non-trivial Chern number of the normal-state bands results necessarily in a finite vorticity of the superconducting order parameter $\Delta(\kk)$, which necessitates it vanishing in at least one point in the Brillouin zone (BZ). We note in passing that the position of this gap node need not lie on the Fermi surface (which can be tuned by doping the graphene), such that the BdG spectrum remains generally gapped everywhere in the BZ. 

\begin{figure}\label{fig: normal to bdg}
    \centering
    \includegraphics[width=0.4\textwidth]{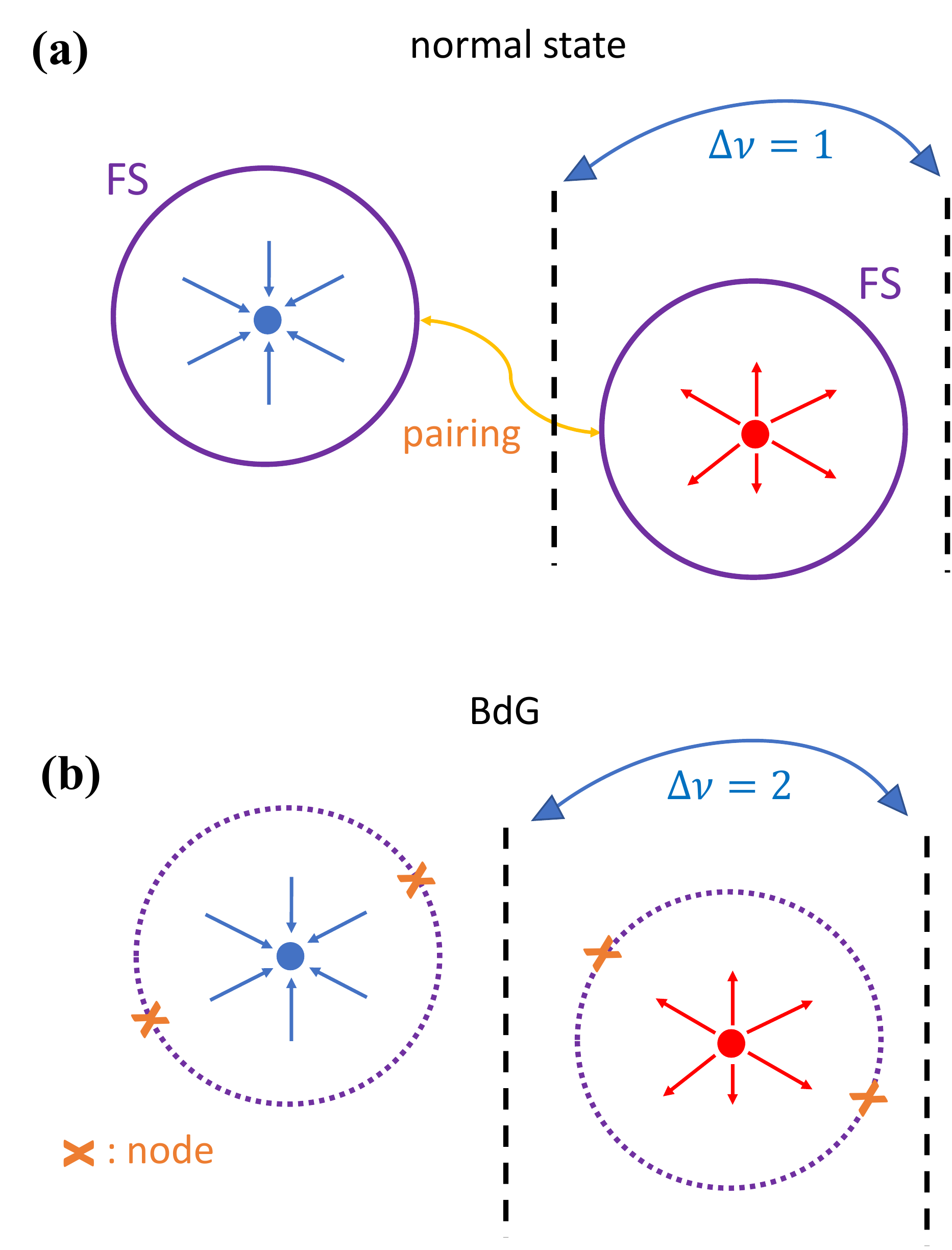}
    \caption{A schematic illustrating how nodes in the normal-state band structure can be inherited in the BdG quasi-particle spectrum of a Weyl or Dirac semimetal. (a) In the normal state 2D (3D) band structure, there is a nonzero change in the winding number (Chern number) across a Dirac (Weyl) point. (b) In the BdG quasi-particle spectrum, there is a nonzero change in the topological invariant (doubled the value of the normal state case) across a Dirac (Weyl) point, guaranteeing the existence of node in the BdG spectrum. Red and blue points denote Dirac (Weyl) points of opposite charge. }
\end{figure}

A similar situation occurs in three-dimensional Weyl semimetals with broken time-reversal symmetry, where the Weyl points serve as the sources and sinks of the Berry curvature in the normal state~\cite{wan-semimetal, armitage2018}. It was shown by Li and Haldane~\cite{li_haldane2018} that a superconducting state formed out of such a semimetal inherits the topology of the normal state, manifest in the fact that the pairing function $\Delta(\kk)$ cannot be defined continuously in the entirely BZ and instead, it must be expanded in terms of monopole harmonics. As a direct consequence, $\Delta(\kk)$ must vanish at least at one point on the Fermi surface, leading to nodes in the BdG spectrum. Similar conclusions regarding the gaplessness of BdG spectra in a Weyl superconductor have also been pointed out in Ref.~\onlinecite{Hosur2014,Meng2012}. 

In this work, we generalize the Li--Haldane result to two spatial dimensions. We provide another perspective on understanding the nodes in the BdG spectrum: the gapless nodes are points of topological transition of the BdG topological invariant. This is illustrated in Fig. \ref{fig: normal to bdg}, in which the topological invariant in the normal state, which lead to protected Fermi surfaces, can be inherited by the BdG Hamiltonian and leads to nodal BdG spectrum. 

We develop a general tool for analyzing the appearance of topologically protected nodes in the BdG spectrum, at least for the cases where the normal state Hamiltonian admits a $\mathbb{Z}$-valued topological invariant. Applying this tool, we reconcile our results with those of Li and Haldane for topological pairing in 3D\cite{li_haldane2018}, and of Murakami and Nagaosa in 2D \cite{murakami_nagaosa2003}. We further extend the discussion of topological pairing to 2D Dirac-type systems, motivated in particular by the superconductivity observed in the twisted bilayer graphene (TBG) near the ``magic" moir\'e twist angle~\cite{TBG-Pablo2018,TBG-Efetov2019}. To this end, we first focus on the general setting of Cooper pairing in graphene-like systems,  combining the analytical and numerical calculations to establish the conditions when the resulting BdG spectrum inherits nodal structure from the Dirac cones found in the normal state. 
 
Our main conclusions regarding the inherited topology of the superconducting state in the twisted bilayer graphene are as follows. We find that for intra-valley intra-sublattice pairing, the BdG spectrum has topologically protected nodes, just like that in the effective Dirac continuum model. For inter-valley inter-sublattice pairing that breaks the $\mathcal C_3$ crystalline symmetry (either broken spontaneously or because of mechanical strain, as observed via scanning tunneling microscopy \cite{choi_nematicity, jiang_nematicity, kerelsky_nematicity} and transport measurements \cite{cao_nematicity}), the BdG spectrum is also necessarily nodal near charge neutrality.

The rest of the paper is organized as follows. In Section \ref{sec: topology}, we provide a general theory of how the existence of a topological invariant of the normal state protects nodes in the BdG spectrum of a superconductor, followed by specific examples of two-dimensional (2D) and three-dimensional (3D) models in the literature. We then apply this theory to the case of the 2D normal state with a chiral symmetry at charge neutrality in section \ref{subsec: Winding number}, and define the notion of the winding number in the Brillouin zone, which is then inherited by the superconductor, resulting in the topologically protected nodes of the BdG spectrum. In Section \ref{subsec: dirac}, we provide a model construction based on Dirac semimetal and consider various pairing scenarios in the continuum theory. We support the above analytical results with the numerical evidence of topologically protected nodes in the BdG spectrum on two examples:
the tight-binding model of monolayer graphene in Section \ref{subsec: MLG} and of twisted bilayer graphene in Section \ref{subsec: TBG}. In both cases, we can understand the presence/absence of node in the BdG spectrum by projecting on the vicinity of the Fermi surface in the Dirac limit. We provide a summary and outlook in Section \ref{sec: conclusion}, focusing in particular on the consequences of our results for twisted bilayer graphene.

\section{Inherited topology}\label{sec: topology}
\subsection{General set-up}\label{subsec: general theory}

To understand the topological origin of point nodes in the BdG spectrum, let us first consider the problem of a {\it gapped} Bloch Hamiltonian $H_{\vec \kappa}$ characterized by a nontrivial topological invariant $\nu$. Slightly more formally, we let $\nu$ be the evaluation map of the topological invariants of Hamiltonians in the given symmetry class, i.e., $\nu(H_{\vec \kappa})$ denotes the topological invariant of $H_{\vec \kappa}$ \footnote{For symmetry classes without a particle-hole or chiral symmetry, the invariant of the Hamiltonian is defined as that for the states below zero energy.}.
For the time being, we do not need to specify the precise nature of the invariant $\nu$: it could be protected by internal and/ or crystalline symmetries; concrete examples will be provided in the subsequent subsections. 

Although the system is an insulator and would not have any natural pairing instability, we can nevertheless insist on adopting a BdG description of the system. We define
\beq 
\label{eq:weak_pairing}
H^{\rm BdG}_{\vec \kappa} = 
\left( 
\begin{array}{cc}
     H_{{\rm e};\vec \kappa} &  \Delta_{\vec \kappa}\\
     \Delta_{\vec \kappa}^\dagger    & - H^*_{{\rm h};-\vec \kappa}
\end{array}
\right),
\eeq
Note that, although we have restricted ourselves to zero-momentum pairing in the above, we have not fully specified the relationship between the electron and the hole parts of the BdG Hamiltonian. For instance, in the case of spin-singlet pairing we can take $H_{{\rm e}; \vec \kappa}$ to be describing spin-up electrons, and $H_{{\rm h}; \vec \kappa}$ that of the spin-down electrons. Alternatively, in systems with strong spin-orbit coupling we could take simply $H_{{\rm e}; \vec \kappa} = H_{{\rm h}; \vec \kappa} = H_{\vec \kappa}$.

We assume $|\Delta_{\vec \kappa}|$ to be much smaller than the gap of the normal state Hamiltonian. In this limit, the BdG Hamiltonian trivially inherits the energy gap of the Bloch Hamiltonian, but what about its nontrivial topology? The (formal) introduction of mean-field superconducting pairing amounts to a symmetry lowering, since the original particle number conservation ${\rm U}(1)$ is reduced to a $\mathbb Z_2$ fermion parity. Other symmetries, like spin rotation and crystalline symmetries, may also be broken by the pairing.
Whether or not the nontrivial topology survives will generally depend on the Altland-Zirnbauer (AZ) symmetry classes involved (both with and without pairing). For instance, if the nontrivial nature of the original Bloch Hamiltonian relies crucially on the ${\rm U}(1)$ charge-conservation symmetry, then $H^{\rm BdG}_{\vec \kappa}$ should be trivialized given the ${\rm U}(1)$ symmetry is broken.

Yet, certain topological invariants are stable against the introduction of superconducting pairing. We will discuss two such examples in the next subsections. For now, however, let us simply suppose that both the original normal-state invariant  $\nu$ and that of the BdG Hamiltonian, $\nu_{\rm BdG}$, are $\mathbb Z$-valued, and that in the zero-pairing limit with $\Delta_{\vec \kappa}\rightarrow 0$ of Eq.~\eqref{eq:weak_pairing} they are related by
\beq \label{eq:bdg_invariant}
\nu_{\rm BdG}(H^{\rm BdG}_{\vec \kappa}) = 
\nu(H_{{\rm e};\vec \kappa}) + 
\nu(- H^*_{{\rm h};-\vec \kappa}).
\eeq
For such problems, $\nu_{\rm BdG}$ is determined by the relationship between $\nu(H_{{\rm e};\vec \kappa})$ and $ \nu(- H^*_{{\rm h};-\vec \kappa})$, namely, whether the electron and hole contributions add up or cancel. In addition, since we assumed a gapped system to start with, \textit{for weak pairing strength the value of $\nu_{\rm BdG}$ is also fixed by that in the zero-pairing limit.}

Next, we consider a smooth family of Bloch Hamiltonians $\{ H_{{\rm e}; \vec \kappa}(t) : t\in[0,1] \}$ with the property $\nu(H_{{\rm e};\vec \kappa}(0)) \neq \nu(H_{{\rm e};\vec \kappa}(1))$, i.e., there is a value $t_* \in (0,1)$ for which $H_{{\rm e};\vec \kappa}(t_*)$ is gapless. 
Correspondingly, we consider another family of Hamiltonians for the hole part with the same properties.
We are interested in the family of BdG Hamiltonian as defined in Eq.~\eqref{eq:weak_pairing}, where the pairing term $\Delta_{\vec \kappa}(t)$ is a smooth function of $t$ and has a magnitude which is much smaller than the gaps at the two limits of $t=0$ and $1$. 
Furthermore, we suppose the symmetry class of the family of BdG Hamiltonian to remain unchanged, i.e., even if the end points at $t=0,1$ may have higher symmetries, we only consider those that are present for all values of $t$.
In general, a nonzero $\Delta_{\vec \kappa}(t_*)$ would lead to a gapped  $H_{\vec \kappa}^{\rm BdG}(t_*)$ even if the normal Hamiltonian  $H_{{\rm e};\vec \kappa}(t_*)$ and $H_{{\rm h};\vec \kappa}(t_*)$ are gapless. However, if the topological difference between the normal state Hamiltonians at $t=0$ and $t=1$ is inherited by the BdG Hamiltonians, i.e., if 
\begin{equation}\label{eq:nodalness}
    \begin{split}
        &\nu(H_{{\rm e};\vec \kappa}(0)) \neq \nu(H_{{\rm e};\vec \kappa}(1))~~\&~~ \nu(H_{{\rm h};\vec \kappa}(0)) \neq \nu(H_{{\rm h};\vec \kappa}(1)) \\
    &~\Longrightarrow\quad
        \nu_{\rm BdG}(H^{\rm BdG}_{\vec \kappa}(0)) \neq \nu_{\rm BdG}(H^{\rm BdG}_{\vec \kappa}(1)),
    \end{split}
\end{equation}
it then follows that there must exist some $t_*' \in (0,1)$ such that $H^{\rm BdG}_{\vec \kappa}(t_*')$ is gapless. Importantly, within our mean-field assumption and for suitable invariants, $\nu_{\rm BdG}(H^{\rm BdG}_{\vec \kappa}(0))$ and $\nu_{\rm BdG}(H^{\rm BdG}_{\vec \kappa}(1))$ could be determined by that of the zero-pairing limit in Eq.~\eqref{eq:bdg_invariant}. 

Clearly, the topological invariants in the zero-pairing limit of a gapped Bloch Hamiltonian depends only on the symmetry class but not on the details of the pairing (since $\Delta_{\vec \kappa} \to 0$ in this limit). In other words, the gaplessness of $H^{\rm BdG}_{\vec \kappa}(t_*')$ is largely independent of the details of the pairing function, and such gaplessness could arise even if one assumes a momentum-independent pairing. These are cases for which an apparently trivial pairing amplitude would nonetheless lead to a nodal superconductor. We will next study two concrete invariants for which such a mechanism is tenable: one is the Chern invariant, relevant to a two-dimensional Fermi surface in three spatial dimensions; the other is the $\mathbb{Z}$-valued winding number invariant, protected by a certain chiral symmetry, which is the key result of this work.

\subsection{Inherited topology in 3D}\label{subsec: Chern number}

Let us first consider the case of the 2D Chern number in a 3D model. To illustrate the idea, it suffices to consider a gapped two-band Bloch Hamiltonian $H_{\vec \kappa}$ in the AZ symmetry class A, which has a $\mathbb Z$-valued invariant: the Chern number $C$ computed on any closed surface of co-dimension 1 in the Brillouin zone.
The following argument readily generalizes to the multi-band case by replacing the single-band Chern number by the multi-band one.
We suppose the chemical potential is set such that one band is filled and the other is empty.
Let $|\psi_{\vec \kappa}^{\pm} \rangle$ be the $\pm$-energy Bloch state of $H_{\vec \kappa}$, i.e., 
\begin{equation}\begin{split}
H_{\vec \kappa} |\psi_{\vec \kappa}^{\pm}\rangle = E_{\vec \kappa}^\pm |\psi_{\vec \kappa}^{\pm} \rangle 
\end{split}\end{equation}
Generally speaking, the two bands have opposite Chern numbers $\mp C$, i.e.,
\begin{equation}\begin{split}
\frac{1}{2\pi}\int  i \left( d  \langle \psi_{\vec \kappa}^{\pm} |\right) \wedge \left( d | \psi_{\vec \kappa}^{\pm} \rangle \right) = \mp C,
\end{split}\end{equation}
where the integral is over a two-dimensional closed surface in the BZ (for concreteness, one could define the integration to be in the $(k_1,k_2,0)$ plane). Using the notations established in the previous subsection, we write $\nu(H_{\vec \kappa}) = C$.

We are interested in the topological invariant of the associated BdG Hamiltonian when we introduce superconducting pairing. Since the Bloch Hamiltonian is assumed to be in class A to start with, which does not have any symmetries aside from the ${\rm U}(1)$ particle number conservation, it is natural that the BdG Hamiltonian will be in the AZ symmetry class D. In two dimensions, class D also has a $\mathbb Z$ invariant, which could be identified simply with the Chern number of the $E<0$ states in the quasi-particle spectrum. In the zero-pairing limit, the $E<0$ states can be identified with those coming from the electron-like and hole-like sub-block of the BdG Hamiltonian Eq.~\eqref{eq:weak_pairing}, and as such Eq.~\eqref{eq:bdg_invariant} holds. It then remains to evaluate the hole contribution to the BdG Chern number. Since
\begin{equation}\begin{split}
-H^{*}_{-\vec \kappa;0}  | \psi^{+}_{-\vec \kappa} \rangle^* =  ( - E^+_{-\vec \kappa})\,| \psi^{+}_{-\vec \kappa} \rangle^*,
\end{split}\end{equation}
where $- E^+_{-\vec \kappa}<0$ and so $ | \psi^{+}_{-\vec \kappa} \rangle^*$ contributes to the total Chern number of the BdG Hamiltonian.
We simply need to note the value of 
\begin{equation}\begin{split}
& \frac{1}{2\pi}\int  i   \left( d    | \psi^{+}_{-\vec \kappa} \rangle^* |\right) \wedge \left( d   | \psi^{+}_{-\vec \kappa} \rangle^* \right)\\
& = 
- \frac{1}{2\pi} \left(\int   i   \left( d  | \psi^{+}_{-\vec \kappa} \rangle |\right) \wedge \left( d  | \psi^{+}_{-\vec \kappa} \rangle \right) \right)^*\\
& =  C.
\end{split}\end{equation}
In other words, the electronic and hole contributions to the total BdG Chern number add up. Schematically, we may write $\nu_{\rm BdG} = 2 \nu$.

This can be intuitively seen by analyzing the chiral edge modes: the Bloch Hamiltonian itself comes with $C$ chiral edge modes which are not trivialized by the introduction of pairing. In the BdG formalism, the quasi-particle chiral modes correspond to chiral Majorana edge modes. Since a complex fermion is formed by two Majorana fermions, the Chern invariant for the BdG Hamiltonian is doubled. 

We can also make connection to the problem of a doped 3D Weyl semimetal studied by Li \& Haldane \cite{li_haldane2018}.
For concreteness, consider a simple two-band model described by the inversion-symmetric Bloch Hamiltonian
\begin{equation}\begin{split}
H_{\vec k} = \sum_{i=1}^2 \sin k_i \sigma_i +  \left(  \sum_{i=1}^3 \cos k_i  - m\right)  \sigma_3 - \mu \sigma_0,
\end{split}\end{equation}
which is a Weyl SM for $ m = 2$. To see why, note that 
\begin{equation}\begin{split}
\sigma_3 H_{\vec k} \sigma_3 = H_{-\vec k},
\end{split}\end{equation}
meaning inversion is represented by $\sigma_3$. Also, by focusing on the eight time-reversal invariant momenta, we see that the lower band has inversion eigenvalue $-1$ at $\Gamma$, and $+1$ everywhere else. This distribution of inversion eigenvalues is known to indicate the existence of Weyl points in the BZ \cite{turner2012,hughes2011}. To verify this claim more explicitly, note that the energy eigenvalues are
\begin{equation}\begin{split}
E_{\vec k} = \mu \pm \sqrt{\sin^2 k_1 + \sin^2 k_2 +\left(  \sum_{i=1}^3 \cos k_i  - m\right)  ^2},
\end{split}\end{equation}
and to see if the gap $\Delta E_{\vec k}$ closes it suffices to focus on momenta for which $\sin k_1 = \sin k_2 = 0$, i.e., for $(k_1,k_2) = (0,0)$, $(\pi,0)$, $(0,\pi)$, $(\pi,\pi)$. Aside from $k_1 = k_2= 0$, we have  $|\cos k_1 + \cos k_2 - m| >1$ with our choice of $m=2$. This means the only possible gap closing happens along the line $(0,0,k_3)$, and, in fact, at $k_3 = \pm \pi/2$. One can further check that the dispersion is linear about these two points, quantifying them as Weyl points.

We may now imagine adding superconductivity to the problem. First, we notice that the 2D slice of normal Hamiltonian $H_{k_1,k_2,0}$ has Chern number $\nu(0) = 1$, and $H_{k_1,k_2,\pi}$ has Chern number $\nu(\pi)=0$. The last momentum $k_3$ plays the role of an interpolation parameter between the two topologically distant limit. Importantly, as shown in Eq.~\eqref{eq:bdg_invariant} the Chern number of the BdG Hamiltonian inherits that of the normal Hamiltonian in the weak-pairing limit, viz.\ $\nu_{\rm BdG} = 2\nu$. As such, Eq.~\eqref{eq:nodalness} holds and the BdG spectrum will be necessarily nodal at some value of $k_3^*\in (0,\pi)$,
even if a momentum-independent on-site pairing is assumed. This is consistent with the analysis in Li--Haldane\cite{li_haldane2018} who argue that the pairing order parameter should be described by monopole harmonics and vanishes at isolated points on the Fermi surface.

\subsection{Murakami--Nagaosa}

The above argument provides another perspective to understand the results first obtained by Murakami and Nagaosa \cite{murakami_nagaosa2003}. Consider the Haldane model with spin singlet pairing. The original perspective is as follows. Define a Berry connection $\vec A^\Delta (\vec k) = -\nabla_{\vec k} \arg \Delta(\vec k)$, where $\Delta(\vec k)$ is the gap function for a BCS pairing term $\Delta(\vec k) a^\dagger_{\vec k} a^\dagger_{-\vec k}$; the resulting Chern number $\nu = \frac{1}{2\pi} \int_{BZ} d^2 \vec k \nabla \times A^\Delta (\vec k)$ is nonzero. The nonzero Chern number for the filled bands carries over to the gap function $\Delta(\vec k)$ such that $\Delta(\vec k)$ also has a nonzero vorticity, guaranteeing the existence of a node in $\Delta(\vec k)$ in the 2D Brillouin zone. 

With our current perspective, we use the chemical potential $\mu$ as a tuning parameter that detects nodes in the BdG spectrum. First, start with a filled Chern band such that the Chern number is nonzero: $\nu(\mu=0) \neq 0$. Then, allow the chemical potential $\mu$ to sweep until all the bands are emptied. In the process, we assume a $\mu$-independent pairing amplitude whose strength is much less than the gap of the Bloch Hamiltonian. 
In the empty limit, the BdG Chern number is clearly zero: $\nu(\mu\to -\infty) = 0$.
By the argument in the preceding subsection, there must be some $\mu_*$ for which the BdG spectrum is gapless. In the single-band, weak-pairing limit we are considering, the nodes of the BdG spectrum arise from the vanishing of the pairing amplitude upon projection onto the Fermi surface states. 
The necessity of such vanishing points in the pairing amplitude somewhere in the 2D Brillouin zone can be reconciled with the Murakami--Nagaosa argument.

\section{Nodal 2D superconductor from inherited topology}\label{sec: 2D inherited topology}
In the preceding section we elaborated on how the BdG Hamiltonian could inherit topology (a nonzero Chern number) from the normal-state Hamiltonian, and how this could lead to a topological obstruction in gapping out the BdG quasi-particle spectrum. In this section, we demonstrate that our discussions around Eqs.~\eqref{eq:bdg_invariant} and~\eqref{eq:nodalness} apply equally well to the 2D case with the one-dimensional winding number playing the role of a topological invariant protected by the chiral symmetry. In the following, we discuss how this could lead to nodal superconductivity starting from the 2D Dirac semimetal.

\subsection{Winding number}\label{subsec: Winding number}

Consider a 2D semimetal. Since we are in one dimension lower compared to the 3D Weyl semimetal analyzed in the previous section, we should replace the 2D Chern number by a 1D invariant, and the winding number protected by the chiral symmetry, which corresponds to the $\mathbb Z$ entry in the ten-fold way for class AIII in 1D, is a possible candidate. 
To this end, let us first consider a normal-state Hamiltonian with a chiral symmetry, i.e., the Bloch Hamiltonian $H (\vec k)$ anticommutes with a chiral symmetry $\Gamma$ at ever momentum $\vec k$. Physically, this could arise from a sublattice symmetry, which could be a good approximate symmetry in certain 2D materials, especially in graphene-based systems near charge neutrality. For now, let us suppose the superconducting pairing respects the sublattice symmetry; we will later discuss how a similar argument applies even when we consider intra-sublattice pairing, in which case the BdG Hamiltonian enjoys a slightly different chiral symmetry.

More concretely, let us consider a basis for which the chiral symmetry $\Gamma$ takes the form
\begin{equation}
\Gamma H_{\vec k} \Gamma = - H_{\vec k};\quad \Gamma = 
\left(
\begin{array}{cc}
\openone & 0\\
0 & -\openone
\end{array}
\right),
\label{eq:chiral-symm}
\end{equation}
where the $\openone$ and $0$ are understood to be square matrices of the appropriate dimensions. In this ``canonical" basis, $H_{\vec k}$ is off-diagonal
\begin{equation}\begin{split}
H_{\vec k} = \left(
\begin{array}{cc}
0 & Q_{\vec k} \\
Q_{\vec k}^\dagger  & 0
\end{array}
\right).
\end{split}\end{equation}

We can consider the 1D winding number defined over any closed loop in the Brillouin zone on which $H_{\vec k}$ remains gapped. Let us compute the invariant along the loop $(k_1, -\pi) \rightarrow (k_1, \pi)$, which is given by \cite{chiu2016}
\begin{equation}\begin{split}
\nu_{k_1} (H_{\vec k})= \frac{1}{2\pi} \int_{-\pi}^{\pi} d {k_2} \, {\rm Tr}\left( Q_{\vec k}^{-1} i \partial_{k_2} Q_{\vec k} \right).
\end{split}\end{equation}
Let us further suppose we have a 2D analog of the 3D Weyl semimetal, i.e., $\nu_{k_1} = 1$ for $|k_1|<\pi/2$, and $\nu_{k_1} = 0$ for $|k_1|>\pi/2$, which implies a gap closing, generically in the form of a Dirac point, at $k_1 = \pi/2$. 

We will apply the same analysis as in Section\ \ref{sec: topology}. First, suppose we pair electrons described by the same Hamiltonian,  $H_{{\rm e}; \vec \kappa} = H_{{\rm h}; \vec \kappa} =  H_{ \vec \kappa}$, say when we consider spin-singlet pairing in a system with spin-rotation invariance.
In such a scenario, we simply replace $Q_{\vec k} \rightarrow - Q_{-\vec k}^*$ in going from the electron to the hole part of the BdG Hamiltonian. 
In the limit of vanishing $\Delta$, it reads
\begin{equation}\begin{split}\label{eq:inter-inter chiral}
H^{\rm BdG} = \left(
\begin{array}{cc}
H_{\vec k} & 0\\
0 & - H_{-\vec k}^*
\end{array}
\right)=
 \left(
\begin{array}{cccc}
0 & Q_{\vec k} & 0 & 0\\
Q_{\vec k}^\dagger & 0 & 0 & 0\\
 0 & 0 & 0 & -Q^*_{-\vec k} \\
 0 & 0 & -Q_{-\vec k}^T & 0 
\end{array}
\right).
\end{split}\end{equation}
We can then evaluate the corresponding winding number for the hole-block, 
\begin{equation}\begin{split}
\nu_{k_1} (-H^*_{-\vec k})=
& \frac{1}{2\pi} \int_{-\pi}^{\pi} d {k_2}\,  {\rm Tr}\left( (Q_{-\vec k}^*)^{-1} i \partial_{k_2} Q_{-\vec k}^* \right)\\
= & \frac{1}{2\pi} \int_{-\pi}^{\pi} d {k_2}\,  {\rm Tr}\left( Q_{-\vec k} ^{-1} (- i \partial_{k_2} )Q_{-\vec k} \right)^*\\
=& \nu_{-k_1} (H_{\vec k}).
\end{split}\end{equation}

Next, we show that Eq.~\eqref{eq:bdg_invariant} holds. Observe that the sublattice symmetry of the BdG Hamiltonian takes the form
\begin{equation}\begin{split}
\tilde \Gamma = \Gamma \oplus \Gamma 
= 
\left(
\begin{array}{cccc}
\openone & 0 & 0 & 0\\
0 & -\openone & 0 & 0 \\
0 & 0 & \openone & 0\\
 0 & 0 & 0 & -\openone  
\end{array}
\right).
\end{split}\end{equation}
To bring $\tilde \Gamma$ back to the canonical form, we interchange the second and third rows and columns. This gives (in the zero-pairing limit)
\begin{equation}\begin{split}
H^{\rm BdG} = \left(
\begin{array}{cccc}
0 & 0 & Q_{\vec k}  & 0\\
 0 & 0 & 0 & -Q^*_{-\vec k} \\
 Q_{\vec k}^\dagger & 0 & 0 & 0\\
 0 &  -Q_{-\vec k}^T & 0 & 0
\end{array}
\right),
\end{split}\end{equation}
and so we see explicitly
\begin{equation}\begin{split}\label{eq:bdg index}
\tilde \nu_{k_1} (H^{\rm BdG}_{\vec k}) =  \nu_{k_1}(H_{\vec k}) + \nu_{-k_1}(H_{\vec k}),
\end{split}\end{equation}
i.e., the topological index of the BdG Hamiltonian becomes twice that of the original normal state. This implies the arguments in Sec.\ \ref{sec: topology} is applicable, so the BdG spectrum is nodal even when the pairing is turned on. We remark that a pairing that breaks the chiral symmetry in the normal state could, generally speaking, trivialize the BdG winding number, resulting in a gapped BdG spectrum. However, we would later see that even with a nonzero chemical potential which breaks the chiral symmetry, the BdG spectrum could still be nodal with the nodes originating from the inherited topology in the chiral-symmetry limit \ref{subsec: TBG}.

In the following section, we will see that the above winding number analysis applies without modification to one of the pairing scenarios in the Dirac Hamiltonian: the inter-valley inter-sublattice pairing. For other forms of pairing, we will need to change the hole invariant $\nu_{-k_1}(H_{\vec k})$ in order to obtain an analogous equation as in Eq.~\eqref{eq:bdg index}; this is explained in Appendix\ \ref{appendix: winding number cases}.

\subsection{Dirac continuum model}\label{subsec: dirac}

\begin{table*}
\caption{\label{table: analytics} Analytical results of BdG Hamiltonian (up to 1st order in the twist angle $\phi$) of a 2D Dirac system with various pairing scenarios. The 4 pairing scenarios considered are a combination of inter-/intra-valley with inter-/intra-sublattice. For the spin degree of freedom, all pairings scenarios assume a spin singlet pairing, such that the BdG relation, $\Delta(k)=-\Delta^T(-k)$, is satisfied after restoring the spin degree of freedom explicitly. $\Lambda_i, \sigma_i$ denotes Pauli matrices in the Nambu and sublattice space respectively. $\vec q = q_1\hat{x}+q_2\hat{y}$ is the momentum measured from the Dirac point.}
\begin{ruledtabular}
\def\arraystretch{1.5}
\begin{tabular}{cccc}
 Pairing & BdG Hamiltonian & Energy eigenvalues & nodal structure\\ \hline
 inter-valley inter-sublattice ($\bar{\text{V}}\bar{\text{S}}$) & 
 \vtop{\hbox{\strut $H=\Lambda_3 (\mu \sigma_0 + vq_1\sigma_1 + vq_2\sigma_2)$}\hbox{\strut \quad \quad $+ \Lambda_0  (\phi v q_2 \sigma_1 - \phi v q_1 \sigma_2) + \Lambda_1  (\Delta \sigma_1)$}} &
 \vtop{\hbox{\strut $E_{\pm}^2 = \mu^2 + (v|\vec q|)^2(1+\phi^2) + \Delta^2$}\hbox{\strut \quad \quad \space $\pm 2v\sqrt{1+\phi^2}\sqrt{(\mu |\vec q|)^2 + (\Delta q_2)^2}$}} & point nodes\\
 
 inter-valley intra-sublattice ($\bar{\text{V}}$S) & 
 \vtop{\hbox{\strut $H=\Lambda_3 (\mu \sigma_0 + vq_1\sigma_1 + vq_2\sigma_2) $}\hbox{\strut \quad \quad $+ \Lambda_0  (\phi v q_2 \sigma_1 - \phi v q_1 \sigma_2) + \Lambda_1  (\Delta \sigma_0)$}} &
 \vtop{\hbox{\strut $E_{\pm}^2 =\mu^2 + (v|\vec q|)^2(1+\phi^2)+ \Delta^2$}\hbox{\strut \quad \quad \space $\pm 2 qv\sqrt{\mu^2(1+\phi^2) + \Delta^2 \phi^2}$}} & gapped\\
 
 intra-valley inter-sublattice (V$\bar{\text{S}}$) & 
 \vtop{\hbox{\strut $H=\Lambda_3 (\mu \sigma_0 + vq_2\sigma_2 + \phi v q_2 \sigma_1) $}\hbox{\strut \quad \quad $+ \Lambda_0  (vq_1\sigma_1 - \phi v q_1 \sigma_2) + \Lambda_1  (\Delta \sigma_1)$}} &
 \vtop{\hbox{\strut $E_{\pm}^2 =\mu^2 + (v|\vec q|)^2(1+\phi^2)+ \Delta^2$}\hbox{\strut \quad \quad \space $\pm 2 qv\sqrt{\mu^2(1+\phi^2) + \Delta^2}$}} & gapped\\
 
 intra-valley intra-sublattice (VS) & 
 \vtop{\hbox{\strut $H=\Lambda_3 (\mu \sigma_0 + vq_2\sigma_2 + \phi v q_2 \sigma_1)$}\hbox{\strut \quad \quad $+ \Lambda_0  (vq_1\sigma_1 - \phi v q_1 \sigma_2) + \Lambda_1  (\Delta \sigma_0)$}} &
 \vtop{\hbox{\strut $E_{\pm}^2 = \mu^2 + (v|\vec q|)^2(1+\phi^2) + \Delta^2$}\hbox{\strut \quad \quad \space $\pm 2v\sqrt{1+\phi^2}\sqrt{(\mu |\vec q|)^2 + (\Delta q_1)^2}$}} & point nodes\\
\end{tabular}
\end{ruledtabular}
\end{table*}

Here, we discuss a simple continuum model, which for some pairing scenarios, exhibits nodal BdG spectra protected by the winding number as analyzed in the previous section. Motivated by pairing in graphene-based systems, we consider a two-dimensional system with valley degree of freedom whose low-energy effective theories at the $K$ and $K'$ valleys are described by a collection of Dirac electrons:

\begin{align}
\label{eq:dirac ham}
     H_{\tau =K} (\vec q, \phi)&= \mu \sigma_0 + v(q_-e^{+i\phi}\sigma_+ + q_+e^{-i\phi}\sigma_-)\\
     H_{\tau =K'} (\vec q, \phi)&= \mu \sigma_0 +v(q_-e^{+i\phi}\sigma_- - q_+e^{-i\phi}\sigma_+),
\end{align}

where $\vec q = (q_1,q_2)^T$ is the momentum measured from a Dirac point, $q_\pm = q_1\pm i q_2$, $\mu$ is the chemical potential, $\sigma_{\pm}=\frac{1}{2}(\sigma_1 \pm i \sigma_2)$ are Pauli matrices in the sublattice space, and $\phi$ denotes an overall twist angle. We introduce $\phi$ in anticipation of the two models we will consider in later sections: for monolayer graphene (in Section \ref{subsec: MLG}), $\phi=0$, whereas for twisted bilayer graphene (in Section \ref{subsec: TBG}), $\phi \approx 1^{\circ}$. The two valleys are related by time-reversal, which is implemented as complex conjugation, such that $H_{\tau =K} (\vec q) = H_{\tau =K'}^* (-\vec q)$.

Next, we suppose there is a momentum-independent superconducting pairing between the Dirac electrons. The system can be described within the mean-field framework by the BdG Hamiltonian 

\beq 
\label{eq:dirac bdg}
H_{\tau\tau'}(\vec q,\phi) = 
\left( 
\begin{array}{cc}
     H_{\tau}(\vec{q},\frac{\phi}{2}) &  \vec\Delta\\
     \vec\Delta^\dagger    & - H^*_{\tau'}(-\vec q,-\frac{\phi}{2})
\end{array}
\right),
\eeq

where $H_{\tau}(\vec{q})$ and $- H^*_{\tau'}(-\vec q)$ are the electron part and the hole part of the BdG Hamiltonian respectively. Due to the valley degree of freedom $\tau$, the BdG Hamiltonian allows for either inter-valley pairing (a zero-momentum pairing between Dirac cones of opposite chiralities):

\begin{equation*}
\label{eq:inter valley}
H_{KK'}(\vec q) = 
\left( 
\begin{array}{cc}
     H_K(\vec q) &  \vec\Delta\\
     \vec\Delta^\dagger    & -H_{K'}(-\vec q)^*
\end{array}
\right),
\end{equation*}

or intra-valley pairing (a momentum-dependent pairing between Dirac cones of the same chirality):

\begin{equation*}
\label{eq:intra valley}
H_{KK}(\vec q) = 
\left( 
\begin{array}{cc}
     H_K(\vec q) &  \vec\Delta\\
     \vec\Delta^\dagger    & -H_K(-\vec q)^*
\end{array}
\right).
\end{equation*}

The pairing block matrix $\vec \Delta$ has a matrix structure due to the sublattice degree of freedom, which allows for either inter-sublattice pairing (implemented as $\vec \Delta = \Delta \sigma_1$) or intra-sublattice pairing (implemented as $\vec \Delta = \Delta \sigma_0$). We have implicitly chosen the \textit{spin singlet} pairing channel in the spin degree of freedom:
\beq
\tilde{\Delta}_{ss'} = (is^y)_{ss'} \vec \Delta \eeq
in order to satisfy the BdG consistency relation $\vec \tilde \Delta(\vec q) = - \vec\tilde \Delta^T (-\vec q)$, where $\tilde \Delta$ is the full pairing block matrix with valley, spin, and sublattice degree of freedom.

The valley and sublattice degree of freedoms give rise to 4 pairing scenarios: inter-valley inter-sublattice ($\bar{\text{V}}\bar{\text{S}}$), inter-valley intra-sublattice ($\bar{\text{V}}$S), intra-valley inter-sublattice (V$\bar{\text{S}}$), and intra-valley intra-sublattice (VS) pairing. From now on, we will refer to the $\bar{\text{V}} \bar{\text{S}}$ and VS pairing as the \textit{diagonal cases}, and the $\bar{\text{V}}$S and V$\bar{\text{S}}$ as the \textit{off-diagonal cases}. This grouping is motivated by the similarity in behaviors within each of these two groups, which will be shown below.

We provide the analytic form of the BdG Hamiltonians, their eigenvalues, and nodal structures for all 4 cases (up to 1st order in the twist angle $\phi$) in Table \ref{table: analytics}; an analytical technique for obtaining the eigenvalues is presented in Appendix \ref{appendix: squaring trick}. In terms of the nodal structure, the diagonal cases exhibits zeros in the eigenvalues, whereas the off-diagonal cases are fully gapped. It is interesting to note that the energy eigenvalues are very similar within the diagonal cases and within the off-diagonal cases. Within the off-diagonal cases, the origin of the gap is different between the $\bar{\text{V}}$S pairing and the V$\bar{\text{S}}$ pairing. For the $\bar{\text{V}}$S pairing, the BdG spectrum is fully gapped even at $\phi=0$. In comparison, the intra-valley inter-sublattice pairing actually exhibits a ring node at $\phi=0$, and a nonzero $\phi$ is required to fully gap out the BdG spectrum.

\begin{figure}\label{fig: dirac}
    \centering
    \includegraphics[width=0.48\textwidth]{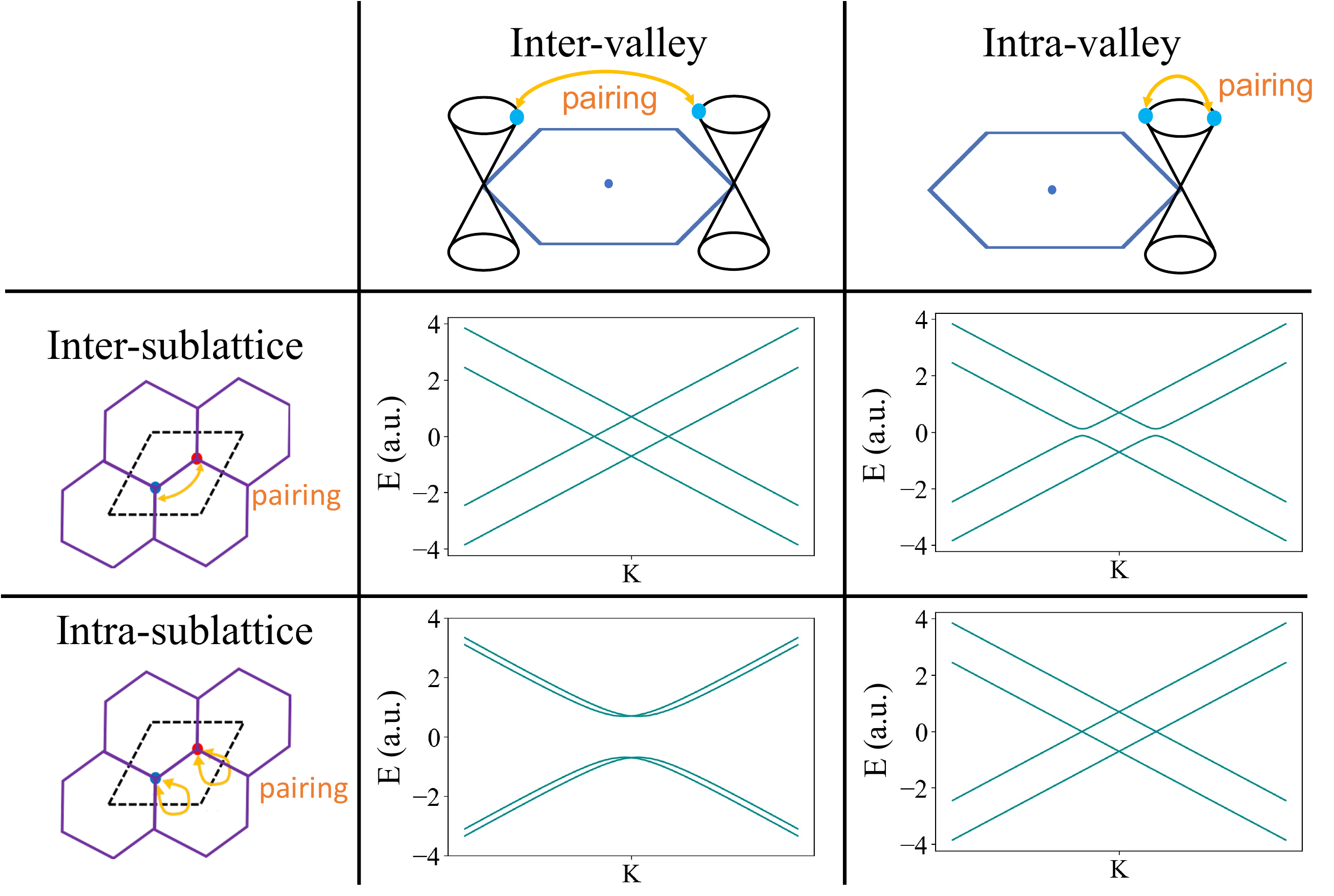}
    \caption{BdG quasi-particle spectra of a 2D Dirac Hamiltonian with various pairing scenarios at pairing parameter $\Delta=0.7$, chemical potential $\mu= 0$, and twist angle $\phi=10^{\circ}$. The 4 pairing scenarios considered are a combination of inter-/intra-valley with inter-/intra-sublattice. The pairing scenarios in the diagonal cases have nodal BdG spectra, with two mini Dirac cones near the $K$ point. The off-diagonal cases are fully gapped.}
\end{figure}

To corroborate the above analytical results, we have also computed the BdG quasi-particle spectra numerically (to all orders in $\phi$), with the results shown in Fig. \ref{fig: dirac}. The most important feature is that the BdG quasi-particle spectrum is nodal (two mini-Dirac cones near the $K$ point) for the diagonal cases, whereas the spectrum is fully gapped for the off-diagonal cases, fully consistent with the analytical results at 1st order of $\phi$. 

Next, we analyze how the momentum-independent pairing acquires a nontrivial pairing symmetry after the projection onto the Fermi surface, focusing on the case of $\bar{\text{V}}\bar{\text{S}}$ pairing.
Assuming the weak-coupling scenario $\Delta \ll \mu$, we Taylor expand one of the energy eigenvalues (at $\phi=0$) to first order in $\Delta/\mu$ to obtain

\begin{equation}\label{eq:energy eigenvalue}
    E^2 \approx (v|\vec q| - \mu)^2 + \Delta^2\left (1 -   \frac{v|\vec q|}{\mu}\sin^2\theta\right),
\end{equation}

where $\theta$ is the azimuthal angle of $\vec q$ with respect to the $k_x$-axis. On the Fermi surface, i.e. when $v|\vec q| = \mu$, Eq.~\eqref{eq:energy eigenvalue} reduces to \mbox{$E= \Delta \cos\theta$}, so we obtain an effective $p$-wave superconductor, with two nodes at $\theta = \pm \pi/2$. The same analysis can be applied to the VS pairing to obtain basically the same result of a $\cos\theta$ angular dependence in the pairing near the Fermi surface. This is reminiscent of the Fu--Kane proximity-induced superconductivity in topological insulator, where even a featureless $s$-wave superconductor results in an effective $p_x + i p_y$ superconductor after projection onto the topologically nontrivial surface states of the topological insulator \cite{fu_kane2008}.

The connection with Fu-Kane can be made exact through a complementary perspective, that is by expressing the pairing in the basis of states on the Fermi surface. This is usually referred to as the \textit{projected pairing} onto the Fermi surface, although it is a bit of a misnomer since there is no projection operator involved; instead, it is just a unitary transformation to the eigen-basis of the normal state Hamiltonian at the Fermi surface. The resulting projected gap function $\mathbf \Delta_p$ for $\bar{\text{V}}\bar{\text{S}}$ pairing is 

\begin{equation}\label{eq: projected pairing diagonal}
    \mathbf\Delta_p = \begin{pmatrix}
\Delta\cos \theta & i \Delta\sin\theta \\
-i\Delta\sin\theta & -\Delta\cos \theta
\end{pmatrix}.
\end{equation}

To first order, the nodal structure of the projected pairing is controlled by the intra-band terms (the diagonal entries), which shows a $\cos\theta$ dependence, consistent with the Taylor series analysis above. For VS pairing, the projected gap function is essentially the same as above (up to an overall phase of $e^{i\theta}$ and some negative signs).

For the off-diagonal ($\bar{\text{V}}\text{S}$ and $\text{V}\bar{\text{S}}$) cases, using the V$\bar{\text{S}}$ pairing as an example, the projected gap function $\mathbf \Delta_p$ is 
\begin{equation} \label{eq: projected pairing off-diagonal}
    \mathbf\Delta_p = \begin{pmatrix}
-\Delta e^{i\theta} & 0 \\
0 & \Delta e^{i\theta}
\end{pmatrix},
\end{equation}

from which it immediately follows that the projected pairing is fully gapped. This is formally exactly the same as the Fu-Kane $p_x + ip_y$ superconductor, with the same property that the projected Hamiltonian respects time reversal symmetry, unlike the conventional spinless $p_x + ip_y$ superconductor.

As noted before, the winding number analysis in Section \ref{subsec: Winding number} applies without modification to the inter-valley inter-sublattice ($\bar{\text{V}}\bar{\text{S}}$) pairing, providing a topological reason for the existence of the point nodes in the BdG spectrum. The details of how the winding number analysis is applied to the other three pairing scenarios ($\bar{\text{V}}$S, V$\bar{\text{S}}$, VS) can be found in Appendix~\ref{appendix: winding number cases}.
The winding number provides a unifying perspective to understand the 4 pairing scenarios considered: the BdG spectrum is nodal or fully gapped, respectively, based on whether the electron and hole winding number (each nonzero from the chirality of the Dirac cone) sums or cancel each other in the BdG Hamiltonian. If the electron and hole winding number sums to give a nonzero BdG winding number (the diagonal cases), the BdG spectra is necessarily nodal, at least in the limit of small pairing strength. On the other hand, if the electron and hole winding number cancel each other out (the off-diagonal cases), the BdG spectrum is gapped.

Although we have started with a momentum-independent pairing term, the resulting BdG quasi-particle spectrum is nodal for certain combinations of chiralities and the matrix structure of the pairing. This can be understood by projecting the pairing onto the Fermi surface, at which point the explicit momentum dependence of the superconducting gap becomes apparent, as shown in Eqs.~\eqref{eq:energy eigenvalue} -- \eqref{eq: projected pairing off-diagonal}.
The point nodes observed in the diagonal cases are inherited from the normal state, i.e. they are necessitated by the topological properties of the normal-state Dirac cones. 

In addition to the above analytical analysis, valid strictly speaking only in the vicinity if the $K$ and $K'$ points (i.e. in the Dirac limit), we have also computed the BdG quasi-particle spectra for various forms of pairing in the realistic tight-binding (TB) models of the monolayer and twisted bilayer graphene. We now turn to the discussion of these results, which fully corroborate the above conclusions.

\section{2D Dirac semimetals}\label{sec: graphene numerics}

Motivated in part by the superconductivity observed in the TBG near the ``magic" moir\'e twist angle~\cite{TBG-Pablo2018,TBG-Efetov2019}, we study whether a superconducting state of TBG may inherit the topology of its normal-state. While the microscopic origin of pairing in TBG is not clear at present, the fact that superconductivity is found near the integer filling fractions, where the ``resets" of the Dirac-like linear density of states are observed~\cite{TBG-resets,TBG-resets2}, motivates us to study superconductivity as arising from a Dirac semimetal. To this end, we will first consider a ``toy" model of $s$-wave pairing in a \textit{monolayer graphene}. In order to make a more realistic analysis, we will then substantiate these conclusions by studying the spin-singlet (s-wave or d-wave) pairing in the tight-binding treatment of TBG. 

As discussed in the previous section, the appearance of the topologically inherited nodes in the BdG spectrum is protected by the 1D winding number, which requires a chiral symmetry in order to be defined rigorously. We note however that the sublattice symmetry used in our preceding analysis in Section \ref{subsec: Winding number} is not exact in either monolayer or twisted bilayer graphene. Nevertheless, it is a good approximate symmetry when we restrict our attention to the states close to the normal-state Dirac points, i.e., for small doping from charge neutrality \cite{chiral_princeton_2021}.
Indeed, we will demonstrate in the following subsections that the signature of topological nodal superconductors persists with the more complex models of monolayer graphene and twisted bilayer graphene. 

\begin{figure*}
    \centering
    \includegraphics[width=0.98\textwidth]{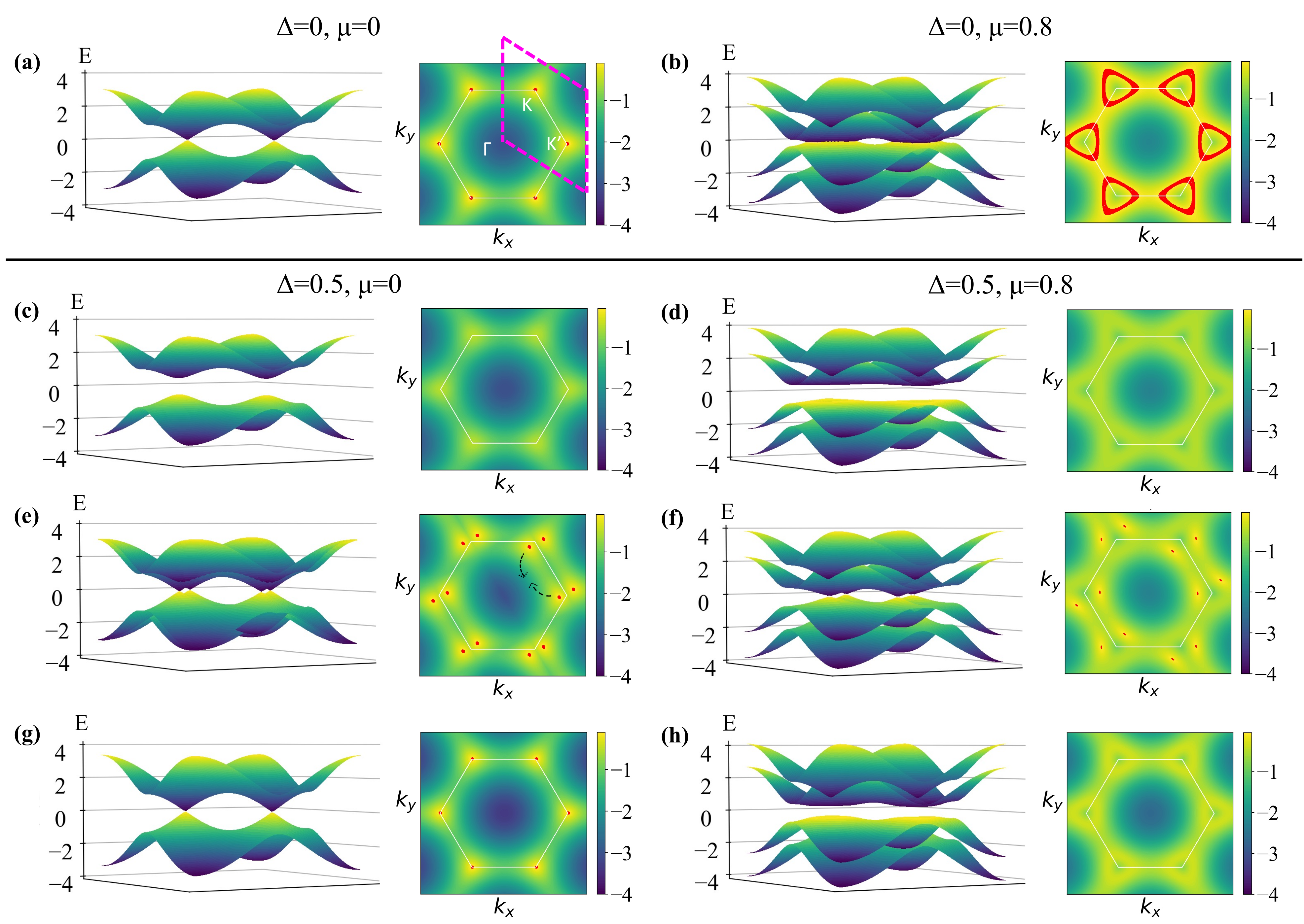}
    \caption{BdG quasi-particle spectra of monolayer graphene with various pairing scenarios: (a)-(b) no pairing, (c)-(d) onsite pairing, (e)-(f) mirror-symmetric inter-sublattice pairing, and (g)-(h) $\mathcal C_3$-symmetric inter-sublattice pairing. The pairing parameter $\Delta$ and chemical potential $\mu$ are: (a) $\Delta=0,\mu=0$; (b) $\Delta=0.5,\mu=0.8$; (c,e,g) $\Delta=0.5,\mu=0$; (d,f,h) $\Delta=0.5,\mu=0.8$. The normal-state is the textbook nearest-neighbor hopping tight-binding model of monolayer graphene. The figures to the right of the 3D plots are contour plots of the lower band over the extended Brillouin zone; red region in the contour plot indicates when the lower band reaches zero energy, i.e. when there is a node. The purple dotted line in the left contour plot of (a) maps the boundary of the region plotted in all the 3D plots. The blacked dotted line in the left contour plot of (e) indicates the movement of the mini-Dirac cones of opposite chiralities as $\mu$ increases, until they meet and gap out the spectrum.}
    \label{fig: MLG}
\end{figure*}

\subsection{Monolayer graphene}\label{subsec: MLG}

We use the nearest-neighbor tight-binding Hamiltonian $\hat{h}$ of monolayer graphene for the normal state. We then add mean-field superconductivity by constructing the BdG Hamiltonian:

\beq 
\label{eq:MLG ham}
H^{\rm BdG}_{\vec \kappa} = 
\left( 
\begin{array}{cc}
     \hat{h}_{\vec \kappa} &  \tilde{\Delta}_{\vec \kappa}\\
     \tilde{\Delta}_{\vec \kappa}^\dagger    & - \hat{h}^*_{-\vec \kappa}
\end{array}
\right).
\eeq
Above, each component is a $2\times 2$ block matrix from the sublattice degree of freedom. 
By construction, we are pairing the states near Dirac cones of opposite chiralities: one at $K$ valley (upper left block matrix) with another at $-K = K'$ valley (lower right block matrix). Therefore, the scope of the monolayer graphene numerical model considered in this section is restricted to inter-valley pairing. For all the cases considered below, the spin pairing channel is always spin singlet $is_2$. 

For the pairing block matrix $\tilde{\Delta}_{\kappa}$, we consider the following four pairing cases: 1) no pairing (just with a redundant BdG degeneracy), 2) intra-sublattice (in particular, the pairing is onsite with a matrix structure of $is_2\otimes \sigma_0$), 3) mirror-symmetric inter-sublattice (with a matrix structure of $is_2\otimes \sigma_1$), and 4) $\mathcal C_3$-symmetric inter-sublattice. For simplicity, we restrict the mirror-symmetric pairing to be only intra-unit-cell, i.e. there is only pairing between each pair of orbitals within the same unit cell. Note that we do not require $\tilde{\Delta}_\kappa$ to have any momentum dependence for the onsite and mirror-symmetric inter-sublattice pairings. 

Numerically, we implement the tight binding model using the PythTB package \cite{pythtb}, and obtain the BdG quasi-particle spectra for the above pairing cases. The nodal structures are consistent with the analytical results for the Dirac Hamiltonians with inter-valley pairing in Section \ref{subsec: dirac}.

Without any pairing, the BdG spectrum [Fig. \ref{fig: MLG}(a)] is just two identical copies of the normal-state band structure at zero chemical potential. At finite chemical potential, the electron part and the hole band of the BdG Hamiltonian shift in opposite directions, forming nodal lines, as shown in Fig. \ref{fig: MLG}(b).

\textbf{Onsite (intra-sublattice, $\bar{\text{V}}\text{S}$) pairing.} For onsite pairing, any nonzero value of $\Delta$ immediately gaps out the BdG spectrum [Fig. \ref{fig: MLG}(c)], and adding a nonzero chemical potential does not close the gap [Fig. \ref{fig: MLG}(d)].

\textbf{Mirror-symmetric inter-sublattice ($\bar{\text{V}}\bar{\text{S}}$) pairing.}

In this case, for small value of $\Delta$ at zero chemical potential, the BdG spectrum exhibits point nodes; each Dirac cone from the normal state splits into two mini-Dirac cones [Fig. \ref{fig: MLG}(e)]. A large $\Delta$ (relative to the hopping parameter) is required to bring these mini-Dirac cones of opposite chirality together to gap out the spectrum. 

Comparing Fig. \ref{fig: MLG}(f) with Fig. \ref{fig: MLG}(b), we see that a nonzero $\Delta$ lifts most of the degeneracy of the nodal lines except at a few points. A nonzero value of $\mu$ breaks the sublattice chiral symmetry, which protects the global existence of point nodes in BdG spectrum according to our winding number analysis in Section \ref{subsec: Winding number}. Nevertheless, we find that the point nodes persist up to some small but finite value of $\mu$ [Fig. \ref{fig: MLG}(f)]. As noted above, this is a manifestation of the fact that the sublattice chiral symmetry, albeit not exact, holds approximately when projected onto the Fermi surface (provided $\mu$ is not too large). In addition, the point nodes in the normal state are protected locally by the $\mathcal C_2T$ symmetry of the monolayer graphene, therefore at a small value of $\mu$, before the nodes move and annihilate each other, there are still point nodes in the BdG spectrum inherited from the normal state. At sufficiently large value of $\mu$ (i.e. far from charge neutrality), these point nodes disappear and the BdG spectrum becomes fully gapped.

\textbf{$\mathcal C_3$-symmetric inter-sublattice ($\bar{\text{V}}\bar{\text{S}}$) pairing.}
In this case, the BdG spectrum is nodal for finite $\Delta$ and $\mu=0$, as seen from Fig. \ref{fig: MLG}(g) and (h). The location of the point node is pinned at the $K$ and $K'$ points because of the additional $\mathcal C_3$ crystalline symmetry. 

Unlike the above mirror-symmetric case, adding any nonzero value of $\mu$ immediately gaps out the BdG spectrum. The reason for this behaviour can be understood as follows: in the normal-state, the Dirac cone is protected by a $\mathbb Z_2$ invariant from the $C_2T$ symmetry. With just $C_2T$ symmetry, the two Dirac cones (regardless of chirality) may merge and gap out the spectrum. When $\mu=0$, the presence of the additional chiral symmetry enhances the topological invariant to a $\mathbb Z$-valued winding number; therefore, two Dirac cones can be localized at the same $k$ point (due to $\mathcal C_3$ symmetry) in the Brillouin zone without gapping out each other. But with nonzero chemical potential $\mu\neq 0$, the chiral symmetry is now broken, and the Dirac cones again become $\mathbb Z_2$ valued and annihilate each other, resulting in a gapped BdG spectrum.

\subsection{Twisted bilayer graphene (TBG)}\label{subsec: TBG}

\begin{figure}
    \centering
    \includegraphics[width=0.47\textwidth]{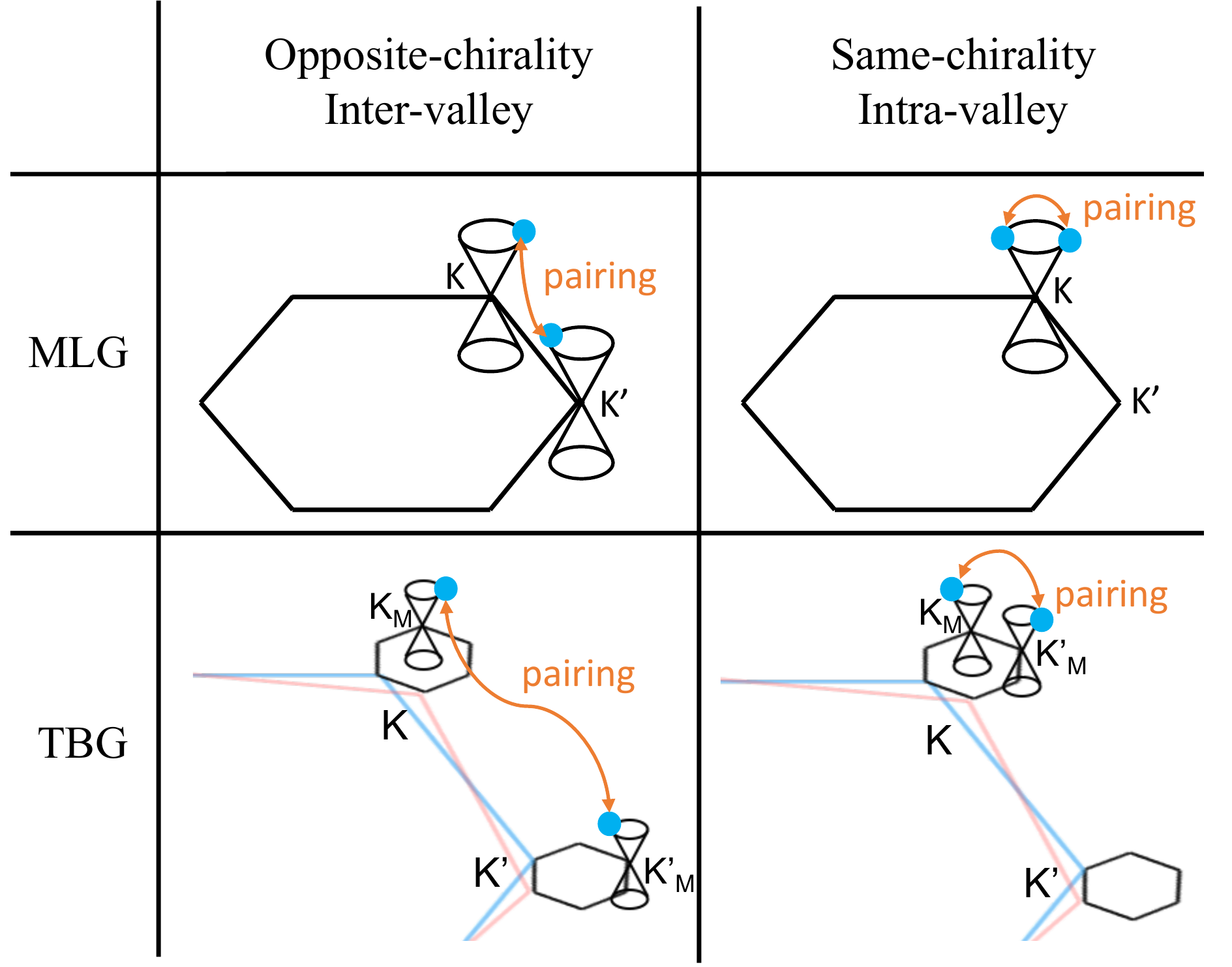}
    \caption{Correspondence between monolayer graphene (MLG) and twisted bilayer graphene (TBG) pairing scenarios. The opposite-chirality pairing in MLG corresponds to inter-valley pairing in TBG; similarly, the same-chirality pairing in MLG corresponds to intra-valley pairing in TBG. $K, K'$ denote the momenta in the microscopic Brillouin zone (before the band folding), and $K_M, K_M'$ denote momenta in the moir\'e Brillouin zone.}
    \label{fig: mono and twisted}
\end{figure}

\begin{figure*}
    \centering
    \includegraphics[width=0.95\textwidth]{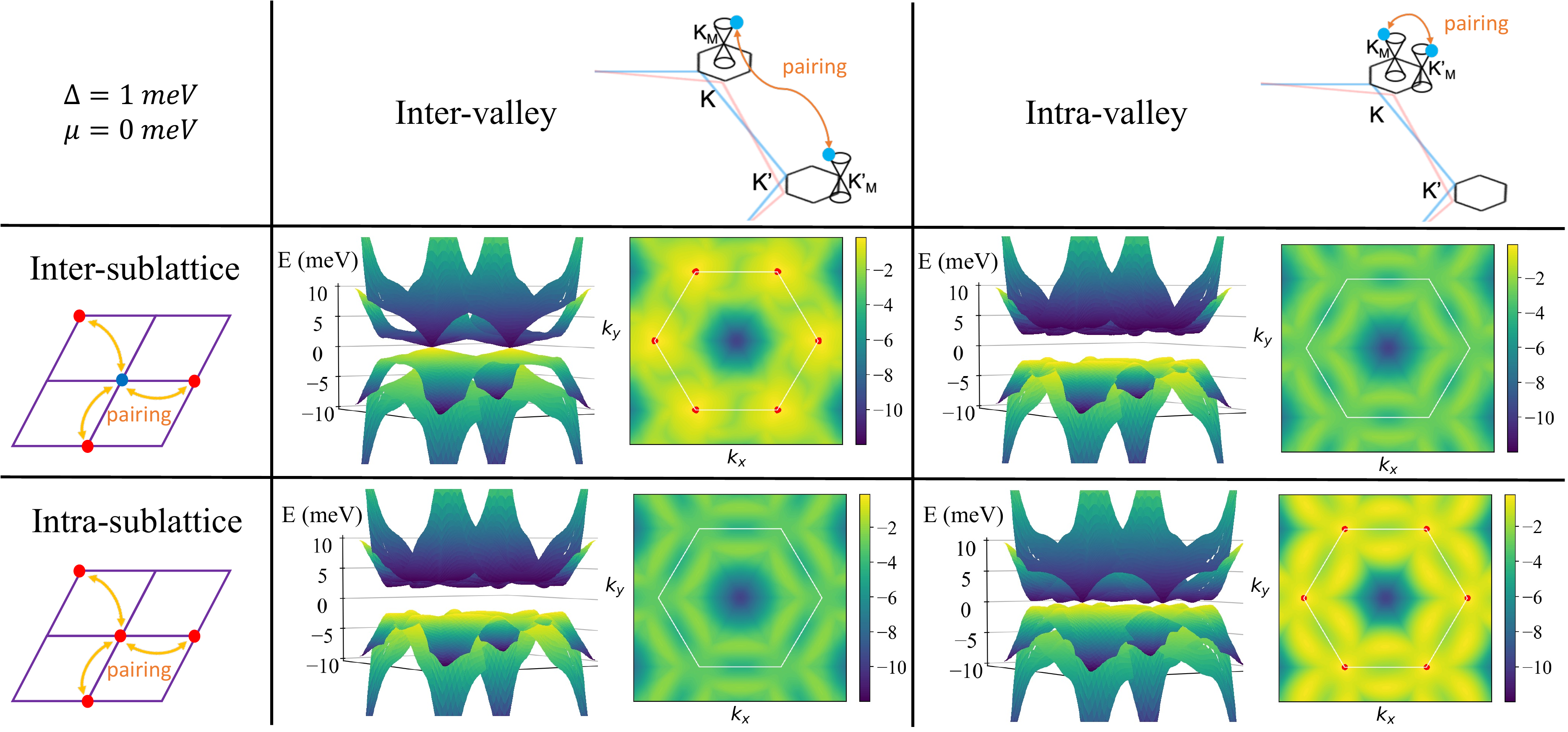}
    \caption{BdG quasi-particle spectra of TBG model at $\Delta=1\, \text{meV}, \mu=0 \,\text{meV}$. The normal-state band structure is taken from the 5-band model of Ref.~\onlinecite{carr2019}. The 4 pairing scenarios considered are the same as that for the Dirac Hamiltonians in Section \ref{subsec: dirac}: either of inter-/intra-valley with either of inter-/intra-sublattice; all pairings are spin singlet, $\mathcal C_3$-symmetric, and inter-moir\'e-unit-cell. Red region in the contour plot indicates the presence of node. The diagonal cases exhibit point nodes at the $K$ and $K'$ points, whereas the off-diagonal cases are completely gapped.}
    \label{fig: tbg C3}
\end{figure*}

Compared to the monolayer graphene model, the magic-angle TBG model has the added ingredients of an explicit valley degeneracy, originating from the microscopic $K$ and $K'$ valleys from the two layers. For monolayer graphene, the Dirac cone at $K$ and $K'$ point of the Brillouin zone have opposite chirality, whereas for TBG, the Dirac cone at $K_M$ and $K_M'$ point of the Moir\'e Brillouin zone have the same chirality \cite{cao_insulator2018,goerbig_montambaux2017,de_gail2011}. Despite these differences, there is still a close correspondence between the pairing scenarios in the monolayer and in TBG, as illustrated in Fig.~\ref{fig: mono and twisted}. The inter-valley (i.e. opposite chirality) pairing in the monolayer graphene is achieved simply by a zero-momentum pairing between $K$ and $K'$ point; for TBG, we need to pair the Dirac cones from different valleys: one Dirac cone in a moir\'e Brillouin zone inherited from the microscopic $K$ point with another Dirac cone inherited from the microscopic $K'$ point.  By the same token, the intra-valley (i.e. same chirality) pairing must necessarily be momentum-dependent in monolayer graphene, however in TBG, the pairing can be moir\'e-momentum independent \footnote{
We note that for intra-valley pairing, the Cooper pairs will transform nontrivially under the microscopic translation. Yet, in the moir{\'e} problem the microscopic translation becomes an internal valley-${\rm U}(1)$ symmetry, which is broken by the intra-valley pairing.}, so the model for same-chirality pairing in TBG can be implemented numerically without requiring a momentum cutoff as one would for a Fulde–Ferrell–Larkin–Ovchinnikov (FFLO) state numerical model. 

There is another nontrivial feature in the TBG model: the normal state of TBG is known to possess a ``fragile topology," manifested in the fact that the set of valley-projected active bands near charge neutrality does not admit a Wannier representation, i.e. cannot be captured by a lattice tight-binding model restricted to localized Wannier orbitals~\cite{po-fragile2019}. Instead one must include higher-lying trivial bands into the lattice tight-binding model~\cite{po-fragile2018}. Therefore, we adopt the effective 5-band (per valley and per spin) tight-binding model for TBG developed in Ref.~\onlinecite{carr2019} as the normal-state Hamiltonian.

Using the normal-state Hamiltonian at a single valley $h$ from Ref. \onlinecite{carr2019}, we include valley degree of freedom by $h\oplus h^*$, where $h^*$ is the time-reversal copy of $h$ at the opposite microscopic valley. We then construct the BdG matrix and impose various momentum-independent pairing scenarios, including both the inter-valley and intra-valley pairing. 

There is a subtlety with regard to the implementation of pairing in the TBG model that is not present in the previous section's monolayer graphene implementation; the details of which is presented in Appendix \ref{appendix: tbg implementation}. The main takeaway is that moir\'e-onsite or moir\'e-intra-unit-cell pairing cannot be $\mathcal C_3$ symmetric. In order to obtain a $\mathcal C_3$-symmetric pairing (regardless of whether it is inter-sublattice or intra-sublattice), one need to pair across different unit cells, i.e. the pairing has to have moir\'e momentum dependence. And there is a nontrivial phase winding in the pairing order parameter $\Delta$ for the diagonal cases due to the $p_+, p_-$ orbital characters of TBG flat bands. 

We numerically solve for the BdG quasi-particle spectra of 4 inter-/intra-valley and inter-/intra-sublattice pairing scenarios (all spin singlet and $\mathcal C_3$-symmetric) at $\Delta=1 \,\text{meV}, \mu = 0 \,\text{meV}$ [Fig. \ref{fig: tbg C3}]. At a small value of $\Delta=1\,\text{meV}$ and $\mu=0\,\text{meV}$, we observe the same nodal structure as that in the Dirac Hamiltonian reported in Section \ref{subsec: dirac}: the diagonal cases contains point nodes near the moir\'e $K$ and $K'$ points, whereas the off-diagonal cases are gapped by any value of $\Delta$ (Fig. \ref{fig: tbg C3}), consistent with the Dirac numerical and analytical results. The point nodes have 4-fold degeneracy from the spin and valley degeneracy.

\begin{figure*}
    \centering
    \includegraphics[width=0.95\textwidth]{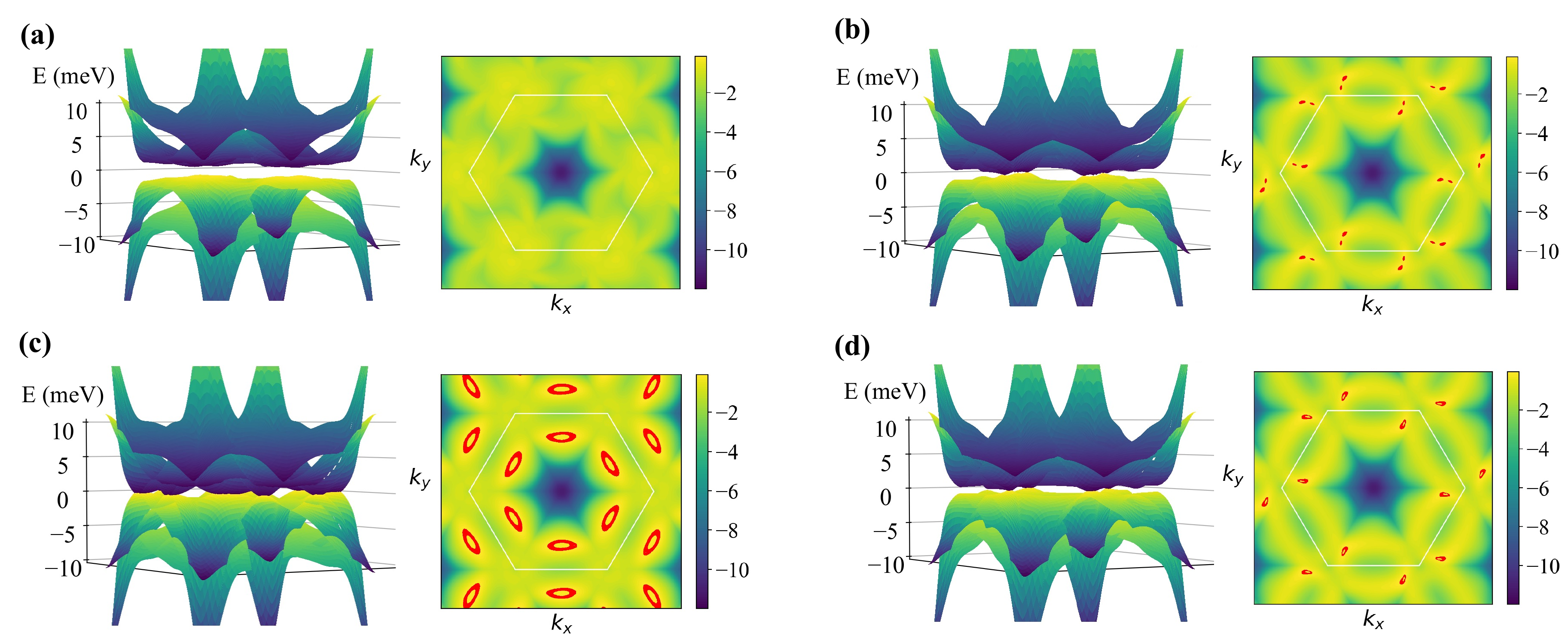}
    \caption{BdG quasi-particle spectra of TBG model at $\Delta = 1 \,\text{meV}, \mu=1.5 \,\text{meV}$. The pairing scenarios considered are: (a) $\mathcal C_3$-symmetric $\bar{\text{V}}\bar{\text{S}}$, (b) mirror-symmetric $\bar{\text{V}}\bar{\text{S}}$, (c) $\mathcal C_3$-symmetric VS and (d) mirror-symmetric VS pairing. Red region in the contour plot indicates the presence of node. Within the Dirac limit (i.e. for small value of $\Delta$ and $\mu$, the point nodes near the $K$ and $K'$ point for the mirror-symmetric cases (b and d) survives up to some small value of $\mu$, whereas point nodes for the $\mathcal C_3$-symmetric cases (a and c) is gapped for any nonzero $\mu$. The ring nodes in the intra-valley cases (c and d) have a different origin than the point node at $K,K'$ points, as explained in Appendix \ref{appendix: ring node}.}
    \label{fig: tbg nonzero mu}
\end{figure*}

Next we consider the effect of adding a nonzero chemical potential and having a $\mathcal C_3$-breaking pairing. Physically, a $\mathcal C_3$-breaking pairing may originate from either spontaneously broken $\mathcal C_3$ symmetry in a nematic superconductor, or from an external perturbation such as strain, both of which are relevant to the experimental measurements on TBG \cite{choi_nematicity, jiang_nematicity, kerelsky_nematicity, cao_nematicity}. We consider both $\mathcal C_3$-symmetric pairing and mirror-symmetric pairing (Fig. \ref{fig: tbg nonzero mu}) at $\Delta=1 \,\text{meV}, \mu = 1.5 \,\text{meV}$. For simplicity, we restrict the mirror-symmetric pairing to only pair within each moir\'e unit cell.

\textbf{$\bar{\text{V}}\bar{\text{S}}$ pairing.} In the inter-valley pairing between the two sublattices, the nodal structure is qualitatively different between the $\mathcal C_3$-symmetric [Fig.\ \ref{fig: tbg nonzero mu} (a)] and the mirror-symmetric counterpart [Fig.\ \ref{fig: tbg nonzero mu} (b)]: any nonzero value of $\mu$ fully gaps out the BdG spectrum for $\mathcal C_3$-symmetric pairing, whereas point node (which splits into 4 distinct nodes at each $K$ and $K'$ point) remains up to some finite value of $\mu$ for mirror-symmetric pairing. This qualitative distinction between the $\mathcal C_3$-symmetric and mirror-symmetric inter-valley pairings matches that observed and explained in the monolayer graphene in Section \ref{subsec: MLG}.

\textbf{VS pairing.} In the case of intra-valley intra-sublattice pairing, both the $\mathcal C_3$-symmetric [Fig.\ \ref{fig: tbg nonzero mu} (c)] and the mirror-symmetric counterpart [Fig.\ \ref{fig: tbg nonzero mu} (d)] exhibit nodal structure. For the $\mathcal C_3$-symmetric case, there remains some 1D ring nodes in the middle of the $\Gamma M$ line. For the mirror-symmetric case, a pair of point nodes for very small value of $\mu$ ``grows'' into a pair of ring nodes as one increases $\mu$. The origin of these 1D ring nodes is distinct from the 0D point node we analyzed in Section \ref{subsec: dirac}; we explain in further details the analysis of these ring nodes for VS pairing in Appendix \ref{appendix: ring node}.

\textbf{``Off-diagonal'' $\bar{\text{V}}$S and V$\bar{\text{S}}$ pairings.} The BdG spectrum remains fully gapped in these cases, as shown in Fig.~\ref{fig: tbg C3}, in full analogy with the pairing in the monolayer graphene discussed in Section \ref{subsec: MLG}.

As a brief connection to the twisted bilayer graphene literature, we remark that if one consider in-plane-phonon-mediated pairing, the $E_2$ phonon mode mediates the $\bar{\text{V}}$S and V$\bar{\text{S}}$ pairings, whereas the $A_1$ phonon mode mediates the $\bar{\text{V}}\bar{\text{S}}$ and $\bar{\text{V}}$S pairing scenarios. For further details, please refer to Appendix \ref{appendix: phonon}.

\section{Conclusion}\label{sec: conclusion}

In this work, we have analyzed how nodal superconductivity could be inherited from a parent 2D Dirac semimetal, similar to the case of a doped 3D Weyl semimetal \cite{li_haldane2018}. Unlike the more conventional cases in which nodal superconductivity arises from finite angular momentum pairing (e.g. $p$- or $d$-waves), our mean-field results apply even when the pairing amplitude is momentum-independent. The presence of nodes in the BdG quasi-particles spectrum depends on the chiralities of the paired Dirac points and whether the pairing is intra- or inter-sublattice; for nodal cases, the existence of point nodes is explained by the inheritance of the normal-state topology -- the winding number that protects the point nodes in the parent nodal semimetal. 

We have analyzed the BdG quasi-particle spectra of the Dirac Hamiltonian analytically and confirmed the nodal structures numerically with tight-binding calculations for the monolayer and twisted bilayer graphene. We have considered four pairing scenarios: the ``diagonal'' cases ($\bar{\text{V}}\bar{\text{S}}$ and VS pairing) and the ``off-diagonal'' cases ($\bar{\text{V}}$S and V$\bar{\text{S}}$ pairing). We find that the BdG spectrum is nodal for the diagonal cases and fully gapped for the off-diagonal cases, provided that the system is sufficiently close to charge neutrality, such that the chiral symmetry (which allows one to define the winding number protecting the point nodes) is a good approximate symmetry. 

We should caution that our numerical results should be interpreted only as a demonstration of the presence of topological nodes and their robustness in the BdG quasi-particle spectrum, rather than the definitive proof of their existence in TBG. While we have based our analysis on the topological properties of the underlying normal-state band structure, no energetic considerations have been invoked. We thus leave a more substantiated discussion on the applicability of our results to the superconductivity in TBG as an open problem.

Nevertheless, it is natural to contemplate on the extent to which our results could be applied to TBG. First of all, although we have restricted our attention to spin-singlet pairing for simplicity, we expect our results to be applicable even to spin-triplet pairing as the analysis relies only on the topological properties of the underlying band structure. However, we also note that our distinction between intra- and inter-valley pairing, motivated by the topological perspective, is different from the more systematic analysis of the pairing symmetries based on the crystallographic point groups. For instance, the $C_3$-breaking pairing we considered would be part of a non-trivial two-component irreducible representation, assuming $C_3$ symmetry is present in the normal state \footnote{We note, though, that strain is known to be significant in typical samples of twisted bilayer graphene, so there could also be extrinsic $C_3$-breaking.}. This perspective is particularly pertinent when the underlying mechanisms for superconductivity are considered, say in the analysis of phonon-mediated pairing. On the other hand, one could also quite reasonably argue that the superconductivity in TBG may come from a strong-coupling mechanism and as such our mean-field analysis does not necessarily apply. 

We should note that at first glance,
the range of electron densities for which superconductivity emerges in experiments on TBG appears to be far from charge neutrality where one expects the Dirac regime to be applicable. 
Yet, the recent experimental results on the ``cascades'' (a.k.a. ``resets'') of Dirac cones~\cite{TBG-resets,TBG-resets2} indicate that the Dirac starting point could be applicable around most, if not all, of the integer fillings. 

In summary, the present work establishes a set of criteria for when the superconductor can inherit the normal-state topology in a two-dimensional Dirac semimetal, similar in spirit to how the 3D monopole superconductivity follows from the topological properties of the Weyl points in the normal state~\cite{li_haldane2018}. Our central result, that the Bogoliubov--de Gennes spectrum is forced to have topologically protected nodes for certain ``diagonal'' combinations of valley and sublattice pairing in the limit of exact chiral symmetry, regardless of microscopic pairing mechanism. In addition, even when chiral symmetry is broken by a finite chemical potential, the nodal nature of the quasi-particle spectrum may survive. As such our results could be applicable to suitable 2D Dirac semimetals, including possibly twisted bilayer graphene. Crucially, we show that the topological nodes in the gap appear  even when the ``bare'' pairing is a momentum-independent  $s$-wave, extending the range of applicability of the present work to quasi-2D Dirac systems with proximitized s-wave pairing.

\begin{acknowledgments}
The authors acknowledge fruitful discussions with Leonid Glazman. 
C.F.B.L. acknowledges support from the Rabi Scholar Program at Columbia University and the Barry M. Goldwater
Scholarship Foundation; H.C.P. was partly supported by the Pappalardo Fellowships at MIT; A.H.N. acknowledges the support of the DOE Basic Energy Sciences (BES) award DE-SC0020177. A.H.N. and H.C.P. thank the hospitality of the Aspen Center for Physics, supported by National Science Foundation grant PHY-1607611, where this project was conceived and a portion of the work performed. 
\end{acknowledgments}

\appendix

\section{Dirac BdG Hamiltonian diagonalization}\label{appendix: squaring trick}

Here, we present analytical techniques for analytically diagonalizing the BdG Hamiltonian for Dirac systems with inter-valley pairing. Note that this is for the case $\phi=0$, where there is no relative twist angle between the layers that contribute to the electron and hole part of the BdG Hamiltonian. 

\subsection{Onsite pairing}

We first present the technique to diagonalize the BdG Hamiltonian for a Dirac Hamiltonian with inter-valley, onsite pairing:
\begin{equation}
H = \Lambda_3\otimes (v\vec q\cdot \vec \sigma +\mu \sigma_0) + \Lambda_1 \otimes (\Delta\sigma_0),
\end{equation}

where $\vec q = q_1\hat{x} + q_x \hat{y}$.

We square the Hamiltonian (which gets rid of cross terms with Nambu matrices $\Lambda_i \neq \Lambda_0$) and obtain
\begin{equation*}
H^2 =  \Lambda_0 \otimes [ (v^2|\vec q|^2+\mu^2+\Delta^2) \sigma_0 +2v \mu (\vec q \cdot \vec\sigma) ].
\end{equation*}

Since we have two (upper left and lower right) identical block matrix, we only need to diagonalize one of them. The remaining diagonalization of the 2 by 2 matrix is straightforward and the energy eigenvalues are
\begin{align}
E^2 &= (v^2|\vec q|^2+\mu^2+|\Delta|^2) \pm 2v \mu|\vec q|\nonumber\\
&=(v|\vec q|\pm \mu)^2 +\Delta^2,
\end{align}
which are always positive for any nonzero pairing parameter $\Delta$, i.e. the BdG quasi-particle spectrum is always gapped.

\subsection{Inter-sublattice pairing}

Here, we diagonalize the BdG Hamiltonian with inter-valley inter-sublattice pairing: 
\beq
H =\Lambda_3 \otimes (v\vec q\cdot \vec \sigma + \mu \sigma_0) + \Lambda_1 \otimes (\vec \Delta \cdot \vec \sigma),
\eeq

where $\vec \Delta = \Delta \hat{x}$.

Using the following identity of Pauli matrices:
\begin{equation}
(\vec{a}\cdot \vec{\sigma})(\vec{b}\cdot \vec{\sigma})=(\vec{a}\cdot \vec{b})I + i(\vec{a}\times \vec{b})\cdot \vec{\sigma}
\end{equation}

and the anti-commutation of Pauli matrices, we square the Hamiltonian to obtain
\begin{align*}
H^2 = \Lambda_0\otimes [ (v^2|\vec q|^2+\mu^2+\Delta^2) \sigma_0 + 2v\mu\vec{q} \cdot \vec{\sigma}]\\
- \Lambda_2 \otimes [2v( \vec{q}\times \vec{\Delta})\cdot \vec{\sigma}].
\end{align*}

We note that the Hamiltonian is now simultaneously diagonalizable with $\Lambda_2$, so we can use eigenstate of $\Lambda_2$ to block diagonalize $H^2$:
\begin{equation*}
H^2_{\pm} =(v^2|\vec q|^2+\mu^2+\Delta^2) \sigma_0 +2v[ \mu \vec q \mp ( \vec{q}\times \vec{\Delta})]\cdot \vec{\sigma},
\end{equation*}

with the $\pm$ sign corresponding to the upper and lower block matrix respectively. Then, directly diagonalize the remaining 2 by 2 block matrix, 
\begin{equation}
E^2 =(v^2|\vec q|^2+\mu^2+\Delta^2) 
\pm 2v\space |\mu \vec{q} \mp (\vec{q}\times \vec{\Delta})|.
\end{equation}
Now we can apply this to the special case $\boldsymbol{\Delta}= \Delta \sigma_1$ considered in the main text:
\begin{equation*}
E^2 =(v^2|\vec q|^2+\mu^2+\Delta^2) \pm 2v \sqrt{(\mu |\vec q|)^2 + (\Delta q_2)^2}.
\end{equation*}

At $\vec q= q_2\hat{y}$, the energy eigenvalues are
\begin{align}
E^2 &=(v^2q_2^2+\mu^2+\Delta^2) \pm2vq_2\sqrt{ \mu^2 +\Delta^2 }\nonumber\\
&= \left(vq_2\pm\sqrt{ \mu^2 +\Delta^2 }\right)^2.
\end{align}

Therefore, for any nonzero value of $\Delta$, the bdG quasi-particle spectrum is always nodal, with two nodes at $(q_1=0, q_2 = \mp\sqrt{ \mu^2 +\Delta^2 }/v)$. 

\subsection{General inter-valley pairing}

Given a general BdG Hamiltonian with inter-valley pairing
\begin{equation}
    H=(v\vec{q} \cdot \vec{\sigma}+\mu \sigma_0)\otimes \Lambda_3 + (\vec{\Delta}_R\cdot \vec{\sigma} )\otimes \Lambda_1-
(\vec{\Delta}_I\cdot \vec{\sigma} )\otimes \Lambda_2,
\end{equation}
where $\vec{ q}=q_1 \hat{x} + q_2 \hat{y}$, and $\vec{\Delta}_R$, $\vec{\Delta}_I$ each has 3 components. The analytical diagonalization of the above Hamiltonian is a nontrivial task. However, with the additional condition that $\vec q, \vec \Delta_R, \vec \Delta_I$ are co-planar, one can analytically diagonalize the Hamiltonian using a "quadratic trick" which we present below.

We follow the same strategy as the previous sections by first squaring the Hamiltonian to obtain 
\begin{align*}
H^2 =&[ (v^2|\vec q|^2+\mu^2+|\Delta|^2) \sigma_0 + 2v\mu\vec{q} \cdot \vec{\sigma}]\otimes \Lambda_0\\
&-2v( \vec{q}\times \vec{\Delta}_I)\cdot \vec{\sigma}\otimes \Lambda_1
-2v(\vec{q}\times \vec{\Delta}_R)\cdot \vec{\sigma}\otimes \Lambda_2\\
&+2v(\vec{\Delta}_R\times \vec{\Delta}_I)\cdot \vec{\sigma}\otimes \Lambda_3,
\end{align*}
where $|\Delta|^2 = |\vec \Delta_R|^2 + |\vec \Delta_I|^2$. However, this Hamiltonian is still not readily diagonalizable analytically. 

To simplify notation, we let 
 $ \xi = v^2|\delta k|^2+\mu^2+|\Delta|^2$ and absorb the factor $2v\vec q \rightarrow \vec q$ such that $H^2$ becomes
\begin{align*}
 H^2 =&[ \xi \sigma_0 + \mu \vec{k} \cdot \vec{\sigma}]\otimes \Lambda_0\\
 &+ (\text{cross terms of $H^2$}),
\end{align*}

where
 \begin{align*}
 (\text{cross terms of $H^2$}) =-(\vec{q}\times \vec \Delta_I)\cdot \vec{\sigma}\otimes \Lambda_1 \\
 -(\vec{q}\times \vec \Delta_R)\cdot \vec{\sigma}\otimes \Lambda_2
+(\vec \Delta_R\times \vec \Delta_I)\cdot \vec{\sigma}\otimes \Lambda_3.
\end{align*}

To begin simplifying the Hamiltonian, we need to add another layer of complexity paradoxically. We square the BdG Hamiltonian again, where we use the condition that $\vec q, \vec \Delta_R, \vec \Delta_I$ are co-planar, such that the fourth power of the BdG Hamiltonian takes on the special form 
\begin{align*}
    H^4 =  [\xi^2+\mu^2|\vec{q}|^2+(\vec{q}\times \vec \Delta_I)^2\\
    +(\vec{q}\times \vec \Delta_R)^2+(\vec \Delta_R\times \vec \Delta_I^2)] \sigma_0 \otimes \Lambda_0\\
+2\xi *\space (\text{cross terms of $H^2$}).
\end{align*}

We observe that the following linear combination between $H^4$ and $H^2$ cancels out the cross terms of $H^2$:
\begin{equation}
    H^4-2\xi H^2 = [\eta \sigma_0- 2\xi \mu (\vec{q} \cdot \vec{\sigma})]\otimes \Lambda_0,
\end{equation}

where we let 

\begin{align*}
   \eta \equiv &-\xi^2+\mu^2|\vec{q}|^2+(\vec{q}\times \vec \Delta_I)^2\\
    &+(\vec{q}\times \vec \Delta_R)^2+(\vec \Delta_R\times \vec \Delta_I^2) 
\end{align*}

to further simplify notation. As the first sign of success, we have successfully block-diagonalized the BdG Hamiltonian at the level of the $\Lambda$ Pauli matrices, so we can focus on just one of the block matrices. The eigenvalues within the $2\times 2$ block matrix are

\begin{equation}
\chi_{\pm} = \eta \pm  2\xi \mu |\vec{q}|.
\end{equation}
Note that $H^4$ and $H^2$ are both simultaneously diagonalized by these eigenvalues.

Importantly, we have found a quadratic equation for which the Hamiltonian (and hence its eigenvalues) satisfies
\begin{equation}
E^4 - 2\xi E^2 - \chi_{\pm} = 0
\end{equation}
which allows us to directly solve for the eigenvalues
\begin{equation}
E^2 = \frac{2\xi \pm \sqrt{4\xi^2 + 4 \chi_{\pm}}}{2}
= \xi \pm \sqrt{\xi^2 + \chi_{\pm}}
\end{equation}

completing the diagonalization.

\section{Variations of winding number analysis}\label{appendix: winding number cases}
\begin{table}
\caption{\label{table: winding} Winding numbers of BdG Hamiltonians of a 2D Dirac system with various pairing scenarios. $\nu_e,\nu_H$ denotes the winding number of the electron-block and hole-block of the BdG Hamiltonian respectively, whereas $\nu_{BdG}$ denote the winding number of the full BdG Hamiltonian. All pairing scenarios exhibit a chiral symmetry at zero chemical potential $\mu$ (in additional to a chiral symmetry from the combination of time-reversal and BdG particle-hole symmetry). $\Lambda_i, \sigma_i$ denotes Pauli matrices in the Nambu and sublattice space respectively. $\vec q = q_1\hat{x}+q_2\hat{y}$ is the momentum measured from the Dirac point.}
\begin{ruledtabular}
\def\arraystretch{1.5}
\begin{tabular}{ccccc}
 Pairing & Chiral symmetry & $\nu_e$ & $\nu_h$ & $\nu_{BdG}$ \\ \hline
 $\bar{\text{V}}\bar{\text{S}}$ & $\Lambda_0\otimes \sigma_3$ & +1 & +1 & +2\\
 $\bar{\text{V}}$S & $\Lambda_3\otimes \sigma_3$ & +1 & -1 &  0\\
 V$\bar{\text{S}}$ & $\Lambda_0\otimes \sigma_3$ & +1 & -1 &  0\\
 VS & $\Lambda_3\otimes \sigma_3$ & +1 & +1 & +2
\end{tabular}
\end{ruledtabular}
\end{table}

In Section \ref{subsec: Winding number}, we show how the winding number from the normal-state Hamiltonian is inherited by the BdG Hamiltonian for the inter-valley inter-sublattice ($\bar{\text{V}}\bar{\text{S}}$) pairing at zero chemical potential. Here, we show that similar arguments apply to the other 3 pairing scenarios ($\bar{\text{V}}$S, V$\bar{\text{S}}$, and VS pairings). The winding numbers are summarized in Table \ref{table: winding}.

\subsection{Inter-valley intra-sublattice $\bar{\text{V}}$S pairing}\label{appendix sub: inter-intra}

In the limit of vanishing $\Delta$, the BdG Hamiltonian with $\bar{\text{V}}$S is the same as that of the $\bar{\text{V}}\bar{\text{S}}$ pairing, which we reproduce below:

\begin{equation}\begin{split}
H^{\rm BdG} = \left(
\begin{array}{cc}
H_{\vec k} & 0\\
0 & - H_{-\vec k}^*
\end{array}
\right)=
 \left(
\begin{array}{cccc}
0 & Q_{\vec k} & 0 & 0\\
Q_{\vec k}^\dagger & 0 & 0 & 0\\
 0 & 0 & 0 & -Q^*_{-\vec k} \\
 0 & 0 & -Q_{-\vec k}^T & 0 
\end{array}
\right).
\end{split}\end{equation}

However, the BdG Hamiltonian manifests a different chiral symmetry

\begin{equation}\begin{split}\label{eq:inter-intra chiral}
\tilde \Gamma 
= 
\left(
\begin{array}{cccc}
\openone & 0 & 0 & 0\\
0 & -\openone & 0 & 0 \\
0 & 0 & -\openone & 0\\
 0 & 0 & 0 & \openone  
\end{array}
\right).
\end{split}\end{equation}

Importantly, the chiral symmetry in the hole block has an extra negative sign, compared to the $\bar{\text{V}}\bar{\text{S}}$ pairing case. Therefore, prior to interchanging the second and third rows and columns, we need to first interchange the third and fourth rows and columns \footnote{This is different from just interchanging the second and the fourth rows and columns, since the order of group operations matters.}. The resultant BdG Hamiltonian is 
\begin{equation}\begin{split}
H^{\rm BdG} = \left(
\begin{array}{cccc}
0 & 0 & Q_{\vec k}  & 0\\
 0 & 0 & 0 & -Q_{-\vec k}^T \\
 Q_{\vec k}^\dagger & 0 & 0 & 0\\
 0 &  -Q^*_{-\vec k} & 0 & 0
\end{array}
\right),
\end{split}\end{equation}

We note that the electron part $Q_{\vec k}$ is same as the $\bar{\text{V}}\bar{\text{S}}$ case, hence the winding number $\nu_{\vec k}(H_{e;k})$ is also the same. On the other hand, the hole part $-Q^T_{-\vec k}$ has an additional adjoint operation, resulting in an additional negative sign in the hole winding number $\nu_{\vec k}(H_{h;-k})$. Therefore, the full BdG winding number is 

\begin{equation}\begin{split}
\tilde \nu_{k_1} (H^{\rm BdG}_{\vec k}) =  \nu_{k_1}(H_{\vec k}) - |\nu_{-k_1}(H_{\vec k})| =0,
\end{split}\end{equation}

since $\nu_{k_1}(H_{\vec k})$ and $\nu_{-k_1}(H_{\vec k})$ have opposite signs. The zero winding number is consistent with the gapped BdG spectrum demonstrated by numerical model in Section \ref{subsec: TBG}.

\subsection{Intra-valley intra-sublattice (VS) pairing}

The chiral symmetry of VS pairing is the same as that given for the case of the $\bar{\text{V}}$S pairing given in Eq.~\eqref{eq:inter-intra chiral}. Nonetheless, the hole Hamiltonian differs by an additional complex conjugate (when compared to the $\bar{\text{V}}$S pairing), which gives rise to an additional negative sign in the hole winding number that cancel with the negative sign from the chiral symmetry as explained in the previous subsection. Ultimately, this makes the hole winding number to be positive, therefore the full BdG winding number is 

\begin{equation}\begin{split}
\tilde \nu_{k_1} (H^{\rm BdG}_{\vec k}) =  \nu_{k_1}(H_{\vec k}) + \nu_{-k_1}(H_{\vec k}) = 2,
\end{split}\end{equation}

where both $\nu_{k_1}(H_{\vec k})$ and $\nu_{-k_1}(H_{\vec k})$ are of the same sign. The nonzero winding number is consistent with the nodal BdG spectrum demonstrated by numerical model in Section \ref{subsec: TBG}.

\subsection{Intra-valley inter-sublattice V$\bar{\text{S}}$ pairing}

The argument from the nodal VS pairing to the gapped V$\bar{\text{S}}$ pairing is in the same vein as that in Appendix \ref{appendix sub: inter-intra}, which is from the nodal $\bar{\text{V}}\bar{\text{S}}$ pairing to the gapped $\bar{\text{V}}$S pairing. The key point is that in both nodal cases, the winding numbers of the electron-block and the hole-block are the same. Then, proceeding to the gapped cases, the hole-block winding number incurs an additional negative sign, thereby trivializing the BdG winding number.

\section{Implementation of TBG pairings}\label{appendix: tbg implementation}

The implementation of TBG inter-sublattice pairing contains subtleties not present in the implementation of monolayer graphene model considered in Section \ref{subsec: MLG}.

First, with respect to the normal state, the 5-band model consists of 2 flat bands and 3 atomic bands, which are needed due to the fragile topological nature of the TBG band structure \cite{po-fragile2018,po-fragile2019}. We use the same convention as that in Ref.~\onlinecite{carr2019}: in the up valley, the 1st band has orbital character $p_+$, and the 2nd band has orbital character $p_-$. The orbital character is reversed in the down valley, where the 1st band has orbital character $p_-$, and the 2nd band has orbital character $p_+$. All flat band orbitals are exponentially localized at the AA sites the moir\'e lattice; we restrict our attention to only pairing between the flat bands. The simplest inter-sublattice term on TBG is a moir\'e onsite term, pairing the $p_+$ to the $p_-$ orbital at the same AA site. And due to the nontrivial winding of the $p_{\pm}$ orbital characters, the moir\'e onsite term is counter-intuitively not $\mathcal C_3$ symmetric.

\begin{figure}
    \centering
    \includegraphics[width=0.47\textwidth]{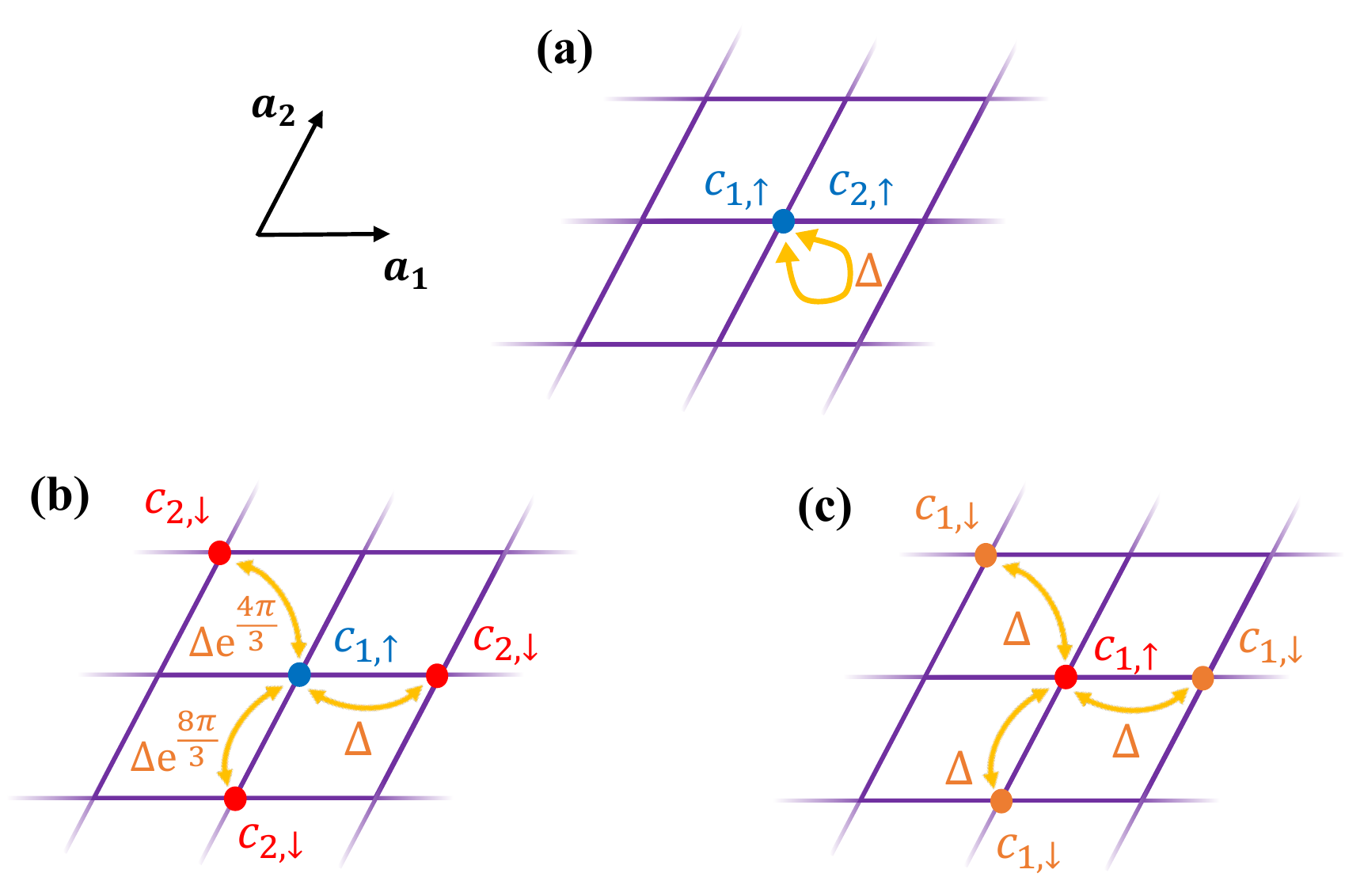}
    \caption{Schematic of the various implementation of pairings in the TBG model: (a) moir\'e-onsite pairing, (b) $\mathcal C_3$-symmetric $\bar{\text{V}}\bar{\text{S}}$ pairing has a phase winding in the pairing $\Delta$,  and (c) $\mathcal C_3$-symmetric $\bar{\text{V}}$S pairing. $c_{1,\uparrow}$ denotes the 1st flat band with mostly $p_+$ orbital character in the up valley, and correspondingly for the other subscripts. $\vec a_1, \vec a_2$ are the two primitive lattice vectors of the triangular lattice formed by the AA sites.}
    \label{fig: tbg pairing}
\end{figure}

There is a qualitative difference between moir\'e-onsite and $\mathcal C_3$-symmetric pairing (inter-moir\'e-unit-cell). Let $\vec a_1,\vec a_2$ denotes the lattice basis vectors for the triangular lattice formed by AA sites, where the flat band orbitals are located, as illustrated in Fig. \ref{fig: tbg pairing}. Moir\'e-onsite pairing [Fig. \ref{fig: tbg pairing}(a)] consists of just one pairing term: for example, for $\bar{\text{V}}\bar{\text{S}}$ pairing, the moir\'e-onsite pairing term is 

\begin{equation*}
    \Delta c^\dagger_{1,\uparrow,\vec 0}c^\dagger_{2,\downarrow, \vec 0} + h.c.,
\end{equation*}

where the subscript $1,2$ denotes the 1st and 2nd flat band respectively, the subscript $\uparrow, \downarrow$ denotes up and down valley respectively, and both orbitals are in the home unit cell. Recalling the aforementioned convention for orbital characters, both $c^\dagger_{1,\uparrow}$ and $c^\dagger_{2,\downarrow}$ has the orbital character of $p_+$. Therefore, under a $2\pi/3$ rotation, both creation operator incurs a phase of $e^{2\pi/3}$, so the onsite pairing term is not $\mathcal C_3$ symmetric. If we insist on making it $\mathcal C_3$ symmetric, then the pairing term vanishes as the coefficient of the pairing term would be the sum of roots of unity $1 + e^{2\pi/3} + e^{4\pi/3}=0$. Therefore, in order to obtain a $\mathcal C_3$-symmetric pairing, the pairing has to be inter-moir\'e-unit-cell. There are two possible scenarios. Case 1: for $\bar{\text{V}}\bar{\text{S}}$ pairing, the pairing terms are

\begin{align*}
    \Delta c^\dagger_{1,\uparrow, \vec 0}c^\dagger_{2,\downarrow, \vec a_1} 
    +\Delta e^{4\pi/3} c^\dagger_{1,\uparrow, \vec 0}c^\dagger_{2,\downarrow, -\vec a_1 + \vec a_2}\\
    +\Delta e^{8\pi/3} c^\dagger_{1,\uparrow, \vec 0}c^\dagger_{2,\downarrow, -\vec a_1 - \vec a_2}  + h.c..
\end{align*}

Since the $\Delta c^\dagger_{1,\uparrow, \vec 0}c^\dagger_{2,\downarrow, \vec a_1}$ term incurs an overall phase of $e^{4\pi/3}$ under a $2\pi/3$ rotation, the 2nd and 3rd term have a phase factor of $e^{4\pi/3}$ and $e^{8\pi/3}$ respectively. Therefore, the pairing has a phase winding, as shown in Fig. \ref{fig: tbg pairing}(b). Similar argument applies to VS pairing, which also has a phase winding in its pairing.

Case 2: for $\bar{\text{V}}$S pairing, the pairing term is of the form 

\begin{equation*}
    \Delta c^\dagger_{1,\uparrow, \vec 0}c^\dagger_{1,\downarrow, \vec a_1} + h.c..
\end{equation*}

Under a $2\pi/3$ rotation, $c^\dagger_{1,\uparrow, \vec 0}$, which has a $p_+$ orbital character, incurs a phase of $e^{2\pi/3}$, whereas $c^\dagger_{1,\downarrow, \vec a_1} $, which has a $p_-$ orbital character, incurs an opposite phase of $e^{-2\pi/3}$. Therefore, the pairing term as a whole does not incur an overall phase, so the pairing parameter $\Delta$ is uniform [Fig. \ref{fig: tbg pairing}(c)]. Similar argument applies to V$\bar{\text{S}}$ pairing, which also has a uniform pairing parameter.

\section{Ring nodes in TBG model with intra-valley intra-sublattice (VS) pairing}\label{appendix: ring node}

\begin{figure}
    \centering
    \includegraphics[width=0.47\textwidth]{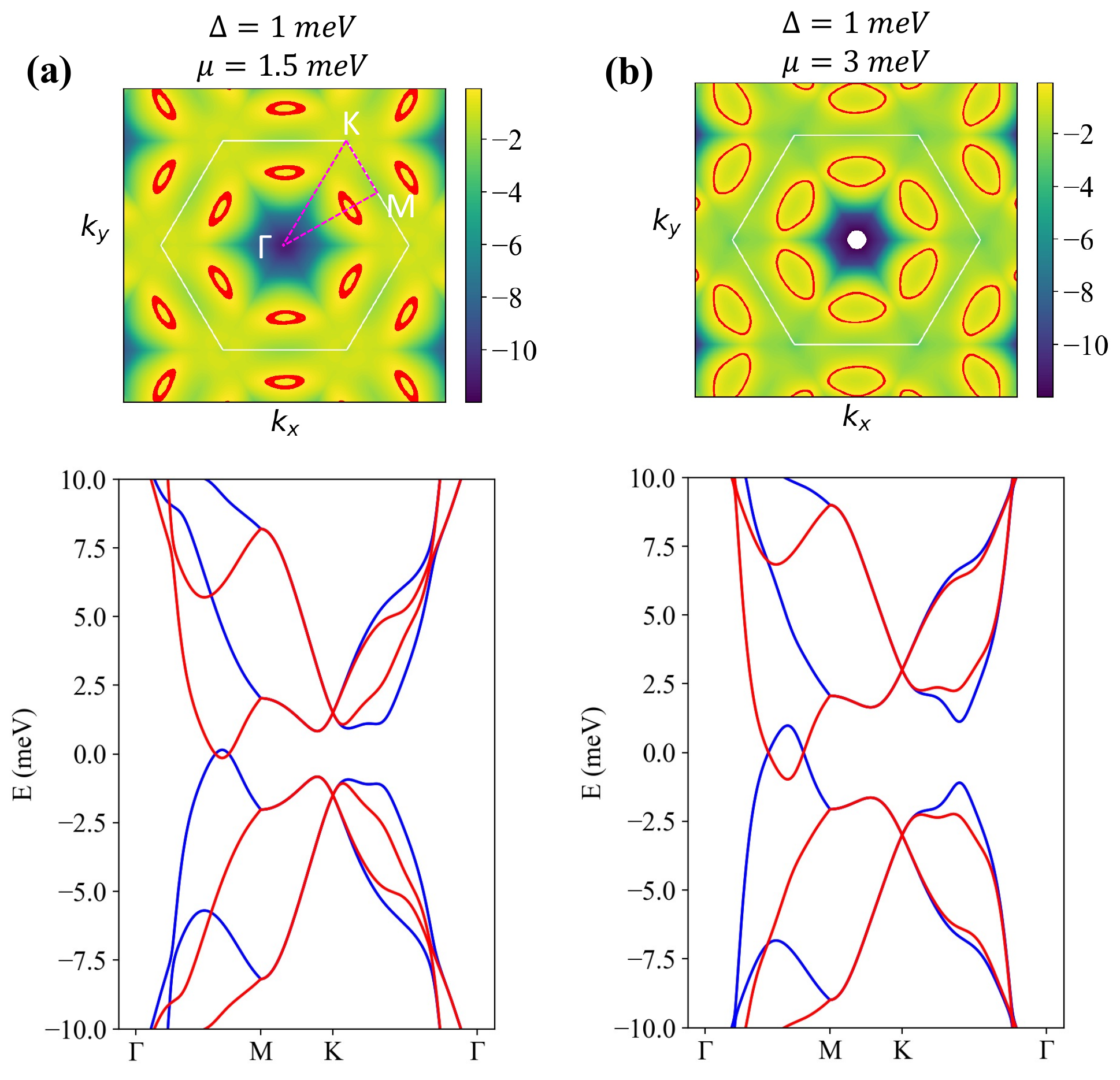}
    \caption{BdG quasi-particle spectra of $\mathcal C_3$-symmetric intra-valley intra-sublattice (VS) pairing in the TBG model at (a) $\Delta = 1 \,\text{meV}, \mu = 1.5 \,\text{meV}$ and (b) $\Delta = 1 \,\text{meV}, \mu = 3 \,\text{meV}$. The upper panels are the contour plots of the lower band closest to zero energy, with red region indicating the presence of ring nodes. The lower panels are the line cuts across the BdG spectra at the high symmetry momenta $\Gamma, M, K, \Gamma$. Blue and red lines corresponds to bands from the up and down valley respectively.}
    \label{fig: ring node}
\end{figure}

For VS pairing (both $\mathcal C_3$-symmetric and mirror-symmetric cases) in the TBG model, the BdG spectrum exhibits ring nodes in the region between the $\Gamma$ and $M$ points (hereafter referred as the $\Gamma-M$ midpoints). This is outside of the scope of the Dirac regime, which only applies to small value of momenta $\vec q$ from the $K,K'$ point. In short, the presence of the ring nodes is not protected by topology; instead, it is due to the band dispersion of the flat bands and the decoupling between the $K$ and $K'$ valleys.

Consider the BdG Hamiltonian for intra-valley intra-sublattice pairing,

\begin{equation}
    H_{BdG} = 
    \begin{pmatrix}
    h &  & \Delta & \\ 
     & h^* &  & \Delta\\ 
    \Delta &  & -h^* & \\ 
     & \Delta &  & -h
    \end{pmatrix},
\end{equation}

where $h$ again denotes the single-valley Hamiltonian adapted from \cite{carr2019}. In the above BdG Hamiltonian, only the Nambu and valley degree of freedom is shown explicitly; $h$ encapsulates matrix structure for the spin and sublattice degree of freedom. We note that the BdG Hamiltonian decouples into two sub-block matrix, one for each valley: 

\begin{equation}
    H_{\uparrow} = 
    \begin{pmatrix}
    h & \Delta \\ 
    \Delta  & -h^* 
    \end{pmatrix}; \quad 
    H_{\downarrow} = 
    \begin{pmatrix}
     h^* &  \Delta\\ 
     \Delta & -h
    \end{pmatrix},
\end{equation}

where $H_{\uparrow}$ and $H_{\downarrow}$ denotes the sub-block matrix corresponding to the up and down valleys respectively. Since the BdG Hamiltonian decouples into block matrices in the valley space, bands from opposite valleys can cross each other without any hybridization/avoided-crossing behaviors that gap out the spectrum.

The flat bands of the normal state TBG near the $\Gamma-M$ midpoint is very close to zero energy. Adding a nonzero $\Delta$ or $\mu$ can push the band across the zero energy and results in the ring nodes. In Fig. \ref{fig: ring node}, the line cut BdG spectrum is plotted with bands from the two valleys illustrated in different colors: when increasing chemical potential from $\mu=1.5\,\text{meV}$ [Fig. \ref{fig: ring node}(a)] to $\mu=3\,\text{meV}$ [Fig. \ref{fig: ring node}(b)], the bands from the two valleys cross over each other (near the $\Gamma-M$ midpoint) without any avoided-crossing, which matches our understanding that the two valleys are decoupled.

\begin{figure}\label{fig: inter_intra_ring}
    \centering
    \includegraphics[width=0.47\textwidth]{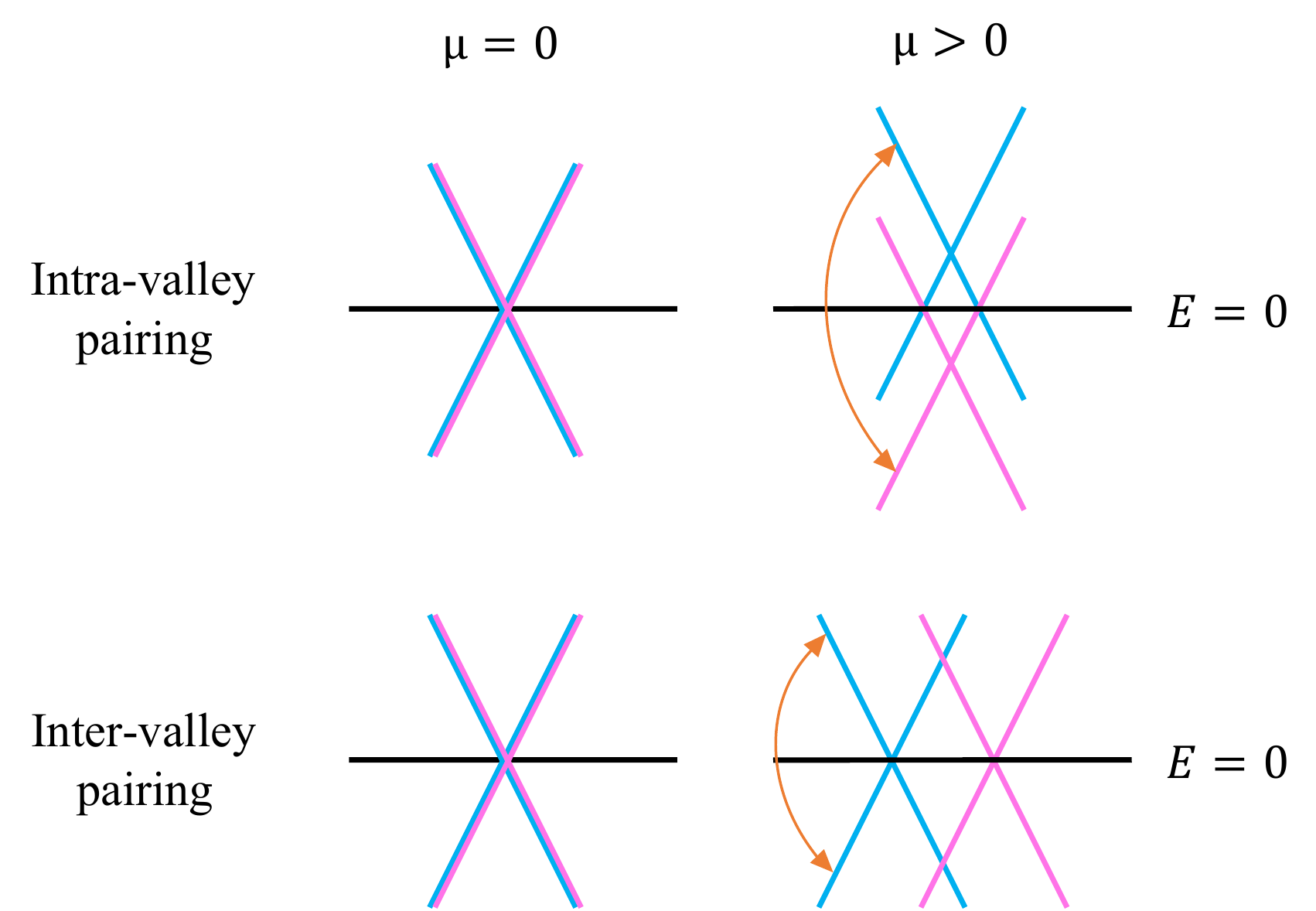}
    \caption{Schematic illustrating the different generic structure of the BdG quasi-particle spectrum between the intra-valley and inter-valley pairing. Orange arrows denotes the BdG chiral symmetry, which relates across (within) valleys for intra-valley (inter-valley) pairing.}
    \label{fig: inter intra ring node}
\end{figure}

We will now explain why ring node only occurs for intra-valley pairings, but not in inter-valley pairings. There is a BdG chiral symmetry by composing the time-reversal symmetry with the BdG particle-hole symmetry/redundancy. As illustrated in Fig.~\ref{fig: inter_intra_ring}, an essential difference between inter- and intra-valley pairings is that the BdG chiral symmetry acts across (within) the valley block structure for intra-valley (inter-valley) pairings. Therefore, with a nonzero chemical potential $\mu$, BdG chiral symmetry pins the Dirac cones to $E=0$ in inter-valley pairings, whereas a Dirac cone in an intra-valley pairing is not necessarily pinned at $E=0$, and may shift up or down (which gives a ring node) as long as its chiral copy shifts in the opposite direction.

\section{Phonon-mediated mechanisms}\label{appendix: phonon}

We base our analysis on Ref.~\onlinecite{wu_macdonald_martin2018}, where superconductivity mediated by in-plane phonon is described by the following interaction Hamiltonian:

\begin{align} \label{eq: phonon interaction Ham}
\begin{split}
    H = -\int d^2 \mathbf r \{g_{E_2}[(\psi^\dagger \tau_3 \sigma_2 \psi)^2 + (\psi^\dagger \tau_0 \sigma_1 \psi)^2 ] \\
    + g_{A_1}[(\psi^\dagger \tau_1 \sigma_1 \psi)^2 + (\psi^\dagger \tau_2 \sigma_1 \psi)^2 ] \},
\end{split}
\end{align}

where $g_{E_2}$ and $g_{A_1}$ are the attractive interaction strength mediated by the $E_2$ and $A_1$ phonon modes respectively, and $\psi = (\psi_{+A},\psi_{+B},\psi_{-A},\psi_{-B})^T$ is a spinor in the valley (labeled by $+,-$) and sublattice (labeled by $A,B$) space.

The inter-valley pairing is energetically more favorable than the intra-valley pairing; therefore, intra-valley pairing are not considered in Ref.~\onlinecite{wu_macdonald_martin2018}. After expanding in terms of the valley and sublattice degree of freedom, and consider only Bardeen-Cooper-Schrieffer (BCS) pairing in the inter-valley channel, the Hamiltonian presented in Ref.~\onlinecite{wu_macdonald_martin2018} is

\begin{equation}
    \begin{aligned}
H=&-4 \int d^{2} \boldsymbol{r}\left\{g_{E_{2}}\left[\hat{\psi}_{+A s}^{\dagger} \hat{\psi}_{-A s^{\prime}}^{\dagger} \hat{\psi}_{-B s^{\prime}} \hat{\psi}_{+B s}+\text { H.c. }\right]\right.\\
&+g_{A_{1}}\left[\hat{\psi}_{+A s}^{\dagger} \hat{\psi}_{-A s^{\prime}}^{\dagger} \hat{\psi}_{+B s^{\prime}} \hat{\psi}_{-B s}+\mathrm{H.c.}\right] \\
&\left.+g_{A_{1}}\left[\hat{\psi}_{+A s}^{\dagger} \hat{\psi}_{-B s^{\prime}}^{\dagger} \hat{\psi}_{+A s^{\prime}} \hat{\psi}_{-B s}+(A \leftrightarrow B)\right]\right\},
\end{aligned}
\end{equation}

where we can identify the 1st term, $\hat{\psi}_{+A s}^{\dagger} \hat{\psi}_{-A s^{\prime}}^{\dagger} \hat{\psi}_{-B s^{\prime}} \hat{\psi}_{+B s}$, as $\bar{\text{V}}$S pairing mediated by the $E_2$ phonon mode; one may also view it as a pair hopping term. The 2nd term, $\hat{\psi}_{+A s}^{\dagger} \hat{\psi}_{-A s^{\prime}}^{\dagger} \hat{\psi}_{+B s^{\prime}} \hat{\psi}_{-B s}$ is also $\bar{\text{V}}$S pairing, but it is mediated by the $A_1$ phonon mode; one may also view it as an inter-sublattice hopping term. The 3rd term, $\hat{\psi}_{+A s}^{\dagger} \hat{\psi}_{-B s^{\prime}}^{\dagger} \hat{\psi}_{+A s^{\prime}} \hat{\psi}_{-B s}$, is $\bar{\text{V}}\bar{\text{S}}$ pairing mediated by the $A_1$ phonon mode.

\begin{table}[b]
\caption{\label{table: phonon mediated pairings} All possible pairing scenarios mediated by the $E_2$ and $A_1$ in-plane phonon modes.}
\begin{ruledtabular}
\def\arraystretch{1.5}
\begin{tabular}{ccccc}
 Pairing & $E_2$ & $A_1$ \\ \hline
 inter-valley inter-sublattice ($\bar{\text{V}}\bar{\text{S}}$) & No & Yes \\
 inter-valley intra-sublattice ($\bar{\text{V}}$S) & Yes & Yes \\
 intra-valley inter-sublattice (V$\bar{\text{S}}$)  & Yes &  No \\
 intra-valley intra-sublattice (VS)  & No & No \\
\end{tabular}
\end{ruledtabular}
\end{table}

On the other hand, we are interested in the question of what are all the possible pairing scenarios generated by various in-plane phonon modes, regardless of energetics. The interactive Hamiltonian given in Eq.~\eqref{eq: phonon interaction Ham} turns out to also give rise to intra-valley pairing term of the form

\begin{equation}
    g_{E_2}[\psi^\dagger_{+A}\psi^\dagger_{+B}\psi_{+A}\psi_{+B} +\mathrm{h.c.}],
\end{equation}

which is a V$\bar{\text{S}}$ pairing mediated by the $E_2$ phonon mode. The $A_1$ phonon mode does not mediate any intra-valley pairing. In addition, neither the $E_2$ or $A_1$ phonon mediate a VS pairing. The full phonon-mediated pairing scenarios are summarized in Table \ref{table: phonon mediated pairings}.

\bibliography{references_updated}
\clearpage

\end{document}